\documentclass{aastex631}

\usepackage{CJK}
\usepackage{amsmath,amsfonts,amssymb}
\usepackage{array,booktabs,ragged2e}
\usepackage{multirow}

\begin{document}
\begin{CJK*}{UTF8}{gbsn}

\title{Mass-loss Rate of Highly Evolved Stars in the Magellanic Clouds}

\author[0009-0002-6102-579X]{Jing Wen (文静)}
\affiliation{Institute for Frontiers in Astronomy and Astrophysics, Beijing Normal University, Beijing 102206, People's Republic of China}
\affiliation{School of Physics and Astronomy, Beijing Normal University, Beijing 100875, People's Republic of China}

\author[0000-0001-8247-4936]{ Ming Yang (杨明) }
\affiliation{Key Laboratory of Space Astronomy and Technology, National Astronomical Observatories, Chinese Academy of Sciences, Beijing 100101, People's Republic of China}

\author[0000-0003-4195-0195]{Jian Gao (高健)}
\affiliation{Institute for Frontiers in Astronomy and Astrophysics, Beijing Normal University, Beijing 102206, People's Republic of China}
\affiliation{School of Physics and Astronomy, Beijing Normal University, Beijing 100875, People's Republic of China}

\correspondingauthor{Ming Yang, Jian Gao}
\email{myang@nao.cas.cn, jiangao@bnu.edu.cn}

\author[0000-0003-2472-4903]{Bingqiu Chen (陈丙秋)}
\affiliation{South-Western Institute for Astronomy Research, Yunnan University, Kunming 650500, People's Republic of China}

\author[0000-0003-1218-8699]{Yi Ren (任逸)}
\affiliation{College of Physics and Electronic Engineering, Qilu Normal University, Jinan 250200, People's Republic of China }

\author[0000-0003-3168-2617]{Biwei Jiang (姜碧沩)}
\affiliation{Institute for Frontiers in Astronomy and Astrophysics, Beijing Normal University, Beijing 102206, People's Republic of China}
\affiliation{School of Physics and Astronomy, Beijing Normal University, Beijing 100875, People's Republic of China}


\begin{abstract}
Asymptotic giant branch stars (AGBs) and red supergiant stars (RSGs) exhibit significant mass loss phenomena and are considered important sources of interstellar dust. In this work, we employed an uniform method of spectral energy distribution fitting to analyze a large, and hence statistically significant, sample of approximately 40,000 RSGs and AGBs in the Magellanic Clouds (MCs), providing a new catalog of evolved stars that includes stellar parameters and dust properties. Our results reveal that the total dust-production rate (DPR) of the Large Magellanic Cloud is approximately $9.69\times10^{-6}\,\rm{M_{\sun }\, yr^{-1}}$, while it is around $1.75\times10^{-6}\,\rm{M_{\sun }\,yr^{-1}}$ for the Small Magellanic Cloud, with a few stars significantly contributing to the total DPR. No significant differences were observed in the contributions to DPR from carbon-rich and oxygen-rich (O-rich) evolved stars in the MCs. We explored the relations between stellar parameters (luminosity, infrared color, period, amplitude) and mass-loss rate (MLR) for evolved stars. A prominent turning point at $\log{(L/L_{\sun})} \approx 4.4$ appears in the luminosity-MLR diagram of RSGs, potentially related to the mass-loss mechanism of RSGs. The luminosity-MLR relation of AGBs is highly scattered. The DPR of AGBs shows a clear change with pulsation period and amplitude, with DPR exhibiting a drastic increase at pulsation periods of approximately 300 days and I-band amplitudes greater than 0.5 mag. Metallicity has some impact on the DPR of O-rich stars, with lower metallicity seeming to result in lower mean DPR and a higher proportion of optically thin stars.

\end{abstract}



\section{Introduction} \label{sec:intro}
The Asymptotic Giant Branch (AGB) is the evolutionary stage near the end of their lives for stars with a mass range of 0.5 to 8 $\rm M_{\sun}$. During the AGB stage, stars produce many elements found in the universe \citep{2014PASA...31...30K}. AGBs lose mass at a high rate, making them major contributors to the cosmic gas and dust cycle, and the rate of mass loss also determines the maximum luminosity they can reach \citep{2018A&ARv..26....1H}. Red supergiant stars (RSGs) play a crucial role in stellar formation and evolution. They are considered the progenitors of Type II-P supernovae (SNe) \citep{2005MNRAS.360..288M, 2009MNRAS.395.1409S}. However, if they undergo substantial mass loss during the RSG phase, some RSGs may evolve to the blue end of the Hertzsprung-Russell diagram, briefly becoming Wolf-Rayet stars or yellow supergiants before further evolving into SNe \citep{2010ASPC..425..247H, 2013EAS....60...31E, 2014ARA&A..52..487S, 2015A&A...575A..60M, 2018MNRAS.475...55B}. The mass loss during the RSG phase has a profound impact on the subsequent evolution of stars. 

The precise mechanism driving mass loss is still unclear, and it may involve a combination of radiation pressure and radial pulsation \citep{2000ApJ...542..120W,2009A&A...498..127V,2014ARA&A..52..487S,2018A&ARv..26....1H}. An increase in luminosity and radiation pressure is expected to accompany a rise in the mass-loss rate (MLR), as supported by observations and model predictions \citep{1975MSRSL...8..369R, 1993ApJ...413..641V, 2005A&A...438..273V}. Additionally, factors such as metallicity and convection may also play a role in influencing mass loss \citep{1992ApJ...397..644M,2001ApJ...551.1073H,2010ApJ...717L..62Y,2011A&A...526A.156M,2016MNRAS.463.1269B}. The true mechanism behind mass loss in AGBs and RSGs remains in debate. Precisely determining the MLR or dust-production rate (DPR) of these evolved stars is necessary to investigate their mass-loss mechanism and the true source of interstellar dust.

Various methods are currently employed to estimate the MLR or DPR of RSGs and AGBs. However, there are significant discrepancies in results among different studies, which may arise from differences in methods, models, and samples \citep{2018A&A...609A.114G,2023A&A...676A..84Y}.   Previous studies have also attempted to derive the relation between MLR and stellar properties such as luminosity, effective temperature, radius, and metallicity \citep{1975MSRSL...8..369R,1988A&AS...72..259D,1992iesh.conf...18F,2017MNRAS.465..403G,1993ApJ...413..641V,1995A&A...297..727B,2002A&A...384..452W,2005A&A...438..273V,2005ApJ...630L..73S,2010A&A...509A..14M,2019A&A...623A.119B}. However, when stars enter the AGB stage, different properties of stars are interrelated and evolve together, making it challenging to isolate the dependence of MLR on individual properties. A simple relation may not fully describe the mass loss of AGBs \citep{2018A&ARv..26....1H,2021ARA&A..59..337D}.
 
In conclusion, despite the progress made in the study of AGBs and RSGs, many issues remain controversial and unclear, awaiting further exploration. Thus, using a unified method to study a large sample of highly evolved stars to obtain statistically significant results is important for our understanding of the complex processes of mass loss and dust production of such stars, which is also beneficial for subsequent research. \citet[hereafter R+21]{2021ApJ...923..232R} constructed samples of RSGs and AGBs in the Local Group. This provides a unique opportunity to study mass loss of evolved stars in the Large Magellanic Cloud (LMC) and Small Magellanic Cloud (SMC) and assess its impact on the interstellar medium and stellar evolution.

The subsequent sections of this paper are structured as follow: Section~\ref{sec:data} provides information about the sample, Section~\ref{sec:fitting} explains the fitting method, Section~\ref{sec:result} displays the findings and discusses the result, and Section~\ref{sec:summary} summarizes the key points.

\section{Data}
\label{sec:data}
The RSG and AGB candidates we used are from the catalog of R+21. This initial sample includes 4,823 RSGs and 31,956 AGBs in the LMC, as well as 2,138 RSGs and 6,004 AGBs in the SMC. They selected the RSGs and AGBs in 12 low-mass galaxies of the Local Group using near-infrared color$-$color diagrams, for which the data were obtained by UKIRT/WFCAM \citep{2004SPIE.5493..411I} and the 2MASS \citep{2006AJ....131.1163S}. For the LMC and SMC, the foreground stars were removed mainly based on the reliable and nearly complete measurements of proper motion and parallax by Gaia EDR3 \citep{2021A&A...649A...1G}. The samples have high completeness and purity. For example, the completeness and purity of RSGs are both over 90\% at a distance modulus ($\mu$) of 18.5 which corresponds to the distance of the LMC \citep{2013Natur.495...76P}.  For detailed information about sample completeness  please refer to Section 5.2.2 in their paper.

To estimate the MLR using the spectral energy distributions (SED) fitting method, it is necessary to construct SEDs covering a wide range of wavelengths. We cross-matched the initial sample with various datasets, including Skymapper DR4 (\textit{u}, \textit{v}, \textit{g}, \textit{i}, \textit{r}, and \textit{z} bands; \citealt{2007PASA...24....1K,2011PASP..123..789B,2018PASA...35...10W,2024arXiv240202015O}), Gaia EDR3 ($G_{\rm{BP}}$, \textit{G}, and $G_{\rm{RP}}$ bands; \citealt{2021A&A...649A...1G}), Magellanic Clouds Photometric Survey (MCPS; \textit{U}, \textit{B}, \textit{V}, and \textit{I} bands; \citealt{2004AJ....128.1606Z}), Vista Magellanic Cloud Survey (VMC; \textit{Y}, \textit{J}, and $K_{\rm{S}}$ bands; \citealt{2011A&A...527A.116C}), InfraRed Survey Facility (IRSF; \textit{J}, \textit{H}, and $K_{\rm{S}}$ bands; \citealt{2007PASJ...59..615K}), Two Micron All Sky Survey (2MASS; \textit{J}, \textit{H}, and $K_{\rm{S}}$ bands; \citealt{2006AJ....131.1163S}), AKARI (\textit{N3}, \textit{N4}, \textit{S7}, \textit{S11}, \textit{L15}, and \textit{L24} bands; \citealt{2007PASJ...59S.401O,2007PASJ...59S.369M,2012AJ....144..179K}), Wide-field Infrared Survey Explorer (WISE; \textit{W1}, \textit{W2}, \textit{W3}, and \textit{W4} bands; \citealt{2010AJ....140.1868W}), and Spitzer ([3.6], [4.5], [5.8], [8.0], and [24] bands; \citealt{2004ApJS..154....1W}), using a cross-matching radius of $1^{\prime \prime}$. Then, we conducted data filtering, as shown in Table~\ref{tab:data}. Our photometric quality selection process aligns with \citealt{Wen_2024} (hereafter Paper \uppercase\expandafter{\romannumeral1}). To distinguish dust types and determine the presence of circumstellar dust, we require that the retained source must have data from [8.0] or $W3$ band and the 2MASS \textit{J}, \textit{H}, and $K_{\rm{S}}$ bands. Despite our data screening efforts, there are still some anomalous $W3$ and $W4$ data. These sources exhibit unusual brightness in  $W3$ and/or $W4$ bands without evident dust features in other bands, resulting in high DPR values. Visual inspection show that, they are typically contaminated by the nearby targets or enviroment due to the low resolution of $W3$ and $W4$ bands. We will label these sources and discuss them in Section~\ref{sec:result}.

As a result, a total of 1,364 sources (1,071 in the LMC and 293 in the SMC) were removed. Among them, 19 sources (17 in the LMC and 2 in the SMC) failed to meet our quality criteria for 2MASS data, and 1,345 sources (1,054 in the LMC and 291 in the SMC) lacked data in $W3$ or [8.0].  Subsequently, we cross-matched our sample with Simbad and removed sources identified as galaxies, planetary nebulae, young stellar object, post-AGB, red giant branch stars, Cepheid variable, RR Lyrae star, blue supergiant, eclipsing binary, ellipsoidal variable, cluster of stars, spectroscopic binary, Orion variable stars etc. There were 57 such sources in the LMC and 24 in the SMC. The final numbers of AGBs and RSGs in the LMC were reduced to 30,937 and 4,714, respectively. In the SMC, the final numbers of AGBs and RSGs became 5,768 and 2,057, respectively. These removed sources, which are based on Simbad ($sflag=1$) and  data quality ($sflag=-1$), will be discussed as factors affecting the total DPR in Section~\ref{sec:result} and included in the machine readable table (MRT) we provided. However, they will not be included in the subsequent analysis. 

Section~\ref{sec:result}

{\catcode`\&=11
\gdef\AandA116{\cite{2011A&A...527A.116C}}}
{\catcode`\&=11
\gdef\2021AandA...649A...1G{\cite{2021A&A...649A...1G}}}
\begin{deluxetable*}{ccccc}
\tabletypesize{\tiny}
\tablecolumns{4}
\renewcommand\arraystretch{1.3}

\tablecaption{Datasets for the spectral energy distribution fitting\label{tab:data}}
\tablehead{\colhead{Datasets}&\colhead{Filters}&\colhead{Selection criteria}&\colhead{LMC N\%$^{\rm a}$}&\colhead{SMC N\%}}
\startdata
 Skymapper$^{\rm b}$ & $u$, $v$, $g$, $i$, $r$, $z$ & $e\_u/v/g/i/r/zPSF<0.1$ &8\%, 12\%, 82\%,  88\%,  93\%,  94\% & 14\%,  21\%,  85\%,  93\%,  93\%\\
MCPS $^{\rm c}$ & \textit{U}, \textit{B}, \textit{V}, \textit{I}, & $e\_U/B/V/Imag<0.1$ &15\% , 69\%, 71\%, 65\% & 30\%, 78\%, 78\%, 69\%\\
 Gaia EDR3 $^{\rm d}$&$G_{\rm{BP}}$, \textit{G}, $G_{\rm{RP}}$&$phot\_bp/g/rp\_mean\_mag\_error < 0.1$&95\%, 99.8\%, 98\%& 95\%, 99.8\%, 97\%\\
VMC$^{\rm e}$&\textit{Y}, \textit{J}, $K_{\rm{S}}$&$e\_Y/J/K_{\rm S}pmag < 0.1$& 28\%, 28\%, 28\% & 57\%, 57\%, 57\%\\
IRSF$^{\rm f}$&\textit{J}, \textit{H}, $K_{\rm{S}}$&$e\_J/H/K_{\rm{S}}mag\_IRSF<0.1$&75\%, 73\%, 65\%&82\%, 80\%, 77\%\\ 
2MASS$^{\rm g}$&\textit{J}, \textit{H}, $K_{\rm{S}}$&\bm{$e\_J/H/K_{\rm S}mag < 0.1$}& 100\%, 100\%, 100\%& 100\%, 100\%, 100\%\\
AKARI$^{\rm h}$& \textit{N3},\textit{N4}, \textit{S7}, \textit{S11}, \textit{L15}, \textit{L24}&$e\_N3/N4/S7/S11/L15/L24mag<0.1$&24\%, 0\%, 22\%, 20\%, 7\% ,22\%&  4\%, 4\%, 3\%, 2\%, 1\%, 0.2\%\\
\multirow{6} *{ WISE$^{\rm i}$}&\multirow{6} *{\textit{W1}, \textit{W2}, \textit{W3}, \textit{W4}}& $e\_W1mag<0.2, e\_W2mag<0.2$& \multirow{6}*{67\%, 65\%, 62\%, 6\%}& \multirow{6}*{ 77\%, 76\%, 54\%, 3\%}\\
{ }& { }&$e\_W3mag<0.3, e\_W4mag<0.3$& \\
{ }&{ }&$\rm{SNR}_\textit{W1}\geq 5, \rm{SNR}_\textit{W2}\geq 5$&\\
{ }&{ }&$\rm{SNR}_\textit{W3}\geq 7, \rm{SNR}_\textit{W4}\geq 10$&\\
{ }&{ }&\bm{$ex\_flag = 0, nb = 1$}&\\
{ }&{ }&$w1/2/3/4cc\_map = 0 $ or $w1/2/3/4flg =0$&\\
Spitzer$^{\rm j}$&[3.6], [4.5], [5.8], [8.0], [24]& $\rm{SNR}\bm{[3.6]/[4.5]/[5.8]/[8.0]}\geq3,closeflag=0$& 84\%, 84\%, 84\%, 84\%, 39\%& 95\%, 96\%, 95\%, 96\%, 31\%
\enddata 
\tablecomments{$^{\rm a}$ The proportion of sources with available data in this band. $^{\rm b}$ 
 \citet{2007PASA...24....1K,2011PASP..123..789B,2018PASA...35...10W} $^{\rm c}$ \citet{2004AJ....128.1606Z} $^{\rm d}$ \2021AandA...649A...1G $^{\rm e}$ 
  \AandA116 $^{\rm f}$ \citet{2007PASJ...59..615K} $^{\rm g}$ \citet{2006AJ....131.1163S} $^{\rm h}$ \citet{2007PASJ...59S.401O,2007PASJ...59S.369M,2012AJ....144..179K} $^{\rm i}$  \citet{2010AJ....140.1868W} $^{\rm j}$ \citet{2004ApJS..154....1W, 2004ApJS..154...25R}}
\end{deluxetable*}
We utilized the extinction map for the LMC and SMC obtained by \citet{2022MNRAS.511.1317C} and applied the extinction law of \citet{1989ApJ...345..245C} to all bands. For the LMC, we adopted $R_{\rm V} = 3.41$, and for the SMC, $R_{\rm V} = 2.74$ \citep{2003ApJ...594..279G}. In the sample, there are 1,337 sources in the LMC and 774 sources in the SMC with $E(B-V) < 0 $ from the extinction maps, possibly due to errors or minimal extinction. For these sources and 11 sources in the LMC not covered by the extinction maps, we applied the method used in the work of \citet{2023A&A...676A..84Y}. We used $R_{\rm V} = 3.1$ and $E(B-V) = 0.1$ for the LMC and $E(B-V) = 0.033$ \citep{2011ApJ...737..103S,2020ApJ...891...57F} for the SMC to correct for foreground extinction and adopted the extinction law derived by \citet{2019ApJ...877..116W}.

\section{Method}
\label{sec:fitting}

One common method to study the MLR of evolved stars is to build SED template libarary based on the dust properties, radiative transfer codes and stellar atmosphere models, then fitting the observed SEDs to the templates. There are many widely used radiative transfer models available now, such as DUSTY \citep{1997MNRAS.287..799I}, 2DUST \citep{2003ApJ...586.1338U}, and MCMax3D \citep{2009A&A...497..155M} and so on. Commonly used central star models include PHOENIX \citep{1995ApJ...445..433A,2013A&A...553A...6H}, ATLAS 9\citep{2003IAUS..210P.A20C} and ATLAS 12 \citep{2005MSAIS...8...14K}, MARCS \citep{2008A&A...486..951G}, and COMARCS \citep{2019MNRAS.487.2133A,2016MNRAS.457.3611A,2009A&A...503..913A} and so on. MARCS covers most of the parameter space of AGBs and RSGs and can be used to construct a unified large template library for evolved stars in the Local Group, including the MCs. Meanwhile, the COMARCS is based on a version of MARCS code with the assumption of spherical symmetry, local thermodynamic and chemical equilibrium. Both MARCS and COMARCS are well suited for purpose of our study with no big difference, which is mostly due to the large uncertainties of mass-loss estimation based on the SED as descripted in Section~5.1 of \citet{2023A&A...676A..84Y}. Our evolved stars SED template library is constructed using the 2-DUST radiative transfer code \citep{2003ApJ...586.1338U}, set to a spherically symmetric model, and adopts the MARCS model, with most parameters consistent with Paper \uppercase\expandafter{\romannumeral1}, and here is a brief introduction.

We generated a new template library containing over 800,000 SEDs based on 868 MARCS models with $[Fe/H]$ ranging from $-1.0$ to 1.0 dex, mass from 0.5 to 15\,$M_{\sun}$, $T_{\rm{eff}}$ from 2500 to 5000\,K, and $\log{g}$ from -0.5 to 0. For oxygen-rich (O-rich) models, we considered 84 values of optical depth $\tau_{\rm 9.7}$ ranging from 0 to 30, while for carbon-rich (C-rich) models, we used 64 $\tau_{\rm 11.3}$ values ranging from 0 to 10. We maintained the settings for dust composition and grain parameters as in Paper \uppercase\expandafter{\romannumeral1}, but expanded the settings for dust shell shape. The silicate optical constants of the O-rich model we adopted come from \citet{1984ApJ...285...89D}, and the amorphous carbon optical constants of the carbon-rich model come from \citet{1996MNRAS.282.1321Z}. For O-rich templates with masses between 0.5 and 5\,$M_{\sun}$, we set the inner radius of the dust shell to eight values: 3, 5, 7, 10, 12, 15, 20, and 25 stellar radii (compared to six values in Paper \uppercase\expandafter{\romannumeral1}), with 20 and 25 only used when $\tau_{\rm 9.7}\geq 10$. For O-rich templates with masses $=$ 15\,$M_{\sun}$, we set the inner radius of the dust shell to 5, 7, 10, 15, 20, 25, 30, and 35, with 30 and 35 stellar radii only used when $\tau_{\rm 11.3}\geq 10$. After generating all over 800,000 templates,  we filtered them based on the dust condensation temperature at the inner radius, which for O-rich models is $<1400$\,K, and for C-rich models is $<1800$\,K \citep{2011A&A...532A..54S,2011ApJ...728...93S}, resulting in 75,492 templates for the LMC （$[Fe/H]=-0.5$ dex）and 72,145 templates for the SMC ($[Fe/H]=-0.75$ dex; \citealt{1997macl.book.....W,1982ApJ...252..461D,2008AJ....136..919B}). The templates and data for the LMC in this study is slightly different from those in Paper \uppercase\expandafter{\romannumeral1}, which has very little impact on the results.

We adopted 50\,kpc for the distance of the LMC \citep{1997macl.book.....W} and 60\,kpc \citep{2011ApJS..192....6L} for the SMC in this work, and then scaled the template flux. We followed the same approach as Paper \uppercase\expandafter{\romannumeral1}, determining the best-fitting model by searching for the minimal $\chi^2$ for each source as,
\begin{equation}
\label{eq:chi}
    \chi^2_i=\sum^N\dfrac{C[\log{F(M_i, \lambda)}-\log{F(O, \lambda)}]^2}{N\left|f(O, \lambda)\right|}
\end{equation}

\begin{figure}
    \centering
	\includegraphics[width=0.9\linewidth]{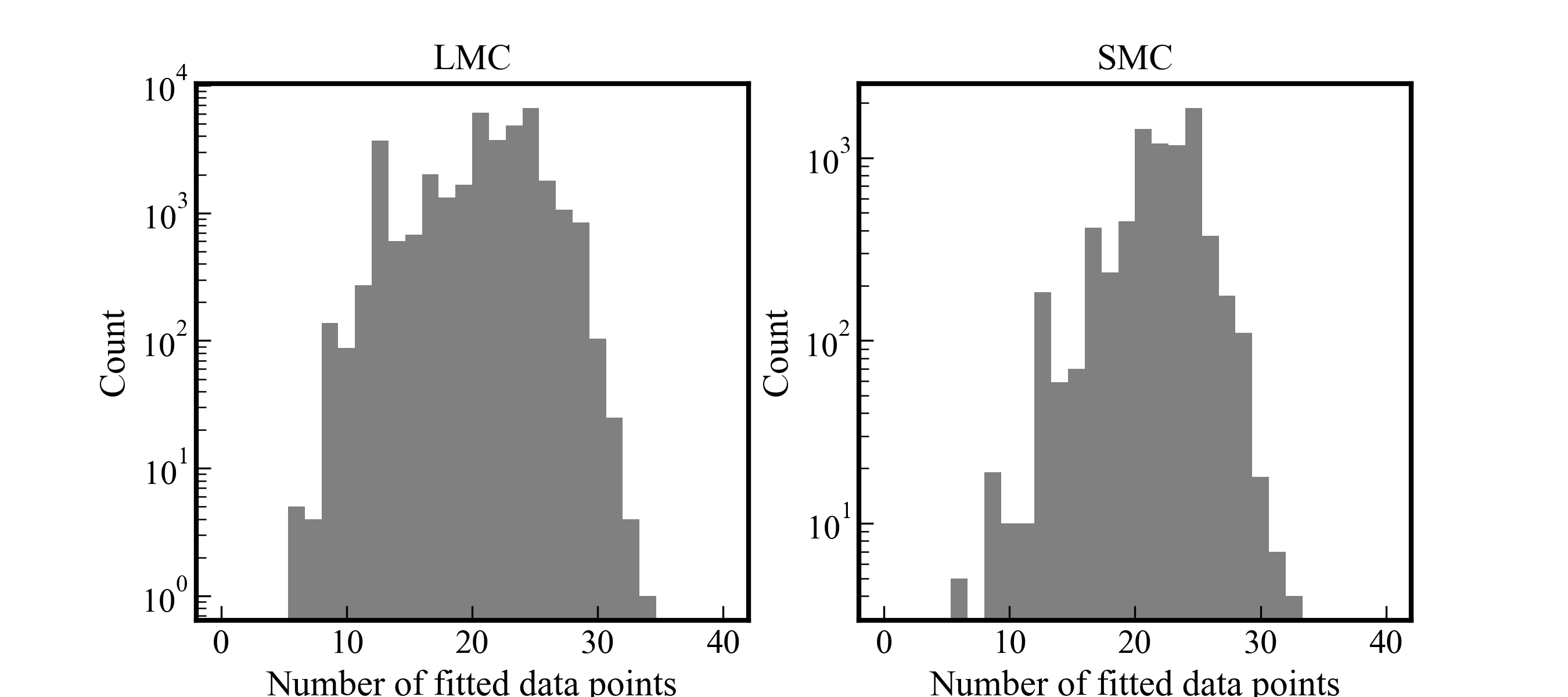}
    \caption{Histogram of numbers of data points for the sample.}
    \label{fig:datan}
\end{figure}

Here, $F(M, \lambda)$ represents the simulated photometric data obtained by convolving the model flux with individual filters\footnote{We utilized the filter response curves from the Spanish Virtual Observatory (SVO) Filter Profile Service \href{http://svo2.cab.inta-csic.es/theory/fps/}{(http://svo2.cab.inta-csic.es/theory/fps/)}}, $F(O, \lambda)$ represents the observed photometric data. $C$ represents the weight applied to data at different wavelengths. For $\lambda \geq 1.0\, \mu m$, we set $C=5$ in order to give more weight to the infrared data. For $\lambda < 1.0\, \mu m$, $C=1$. $N\left|f(O, \lambda)\right|$ is the number of observed data. Each source in our sample has at least 6 bands of data, with a maximum of 34 bands, as depicted in Figure~\ref{fig:datan}.

We divided our samples into two groups for the fitting process. The first group, referred to as S1, was composed of sources lacking photometric data in the $W3$ band or longer wavelengths (6917 in the LMC, 3224 in the SMC). The second group, denoted as S2, included sources with photometric data containing $W3$ or [24], $S11$, $L24$ bands. For the S1 sample, fitting was constrained to models with optical depth $\tau_{\rm 9.7/11.3} \leq 0.01$, while S2 samples were fitted with all models. This grouping helped prevent sources without apparent infrared emission from being inaccurately fitted to templates with high DPR (see Figure 4 of  Paper \uppercase\expandafter{\romannumeral1} for more details). Subsequently, we visually inspected the SEDs of all sources, moved sources that exhibited an obvious rise in the [3.6], [4.5], [5.8], and [8.0] bands from S1 to S2 (292 in the LMC and 65 in the SMC).

\begin{figure*}
    \centering
	\includegraphics[width=1\linewidth]{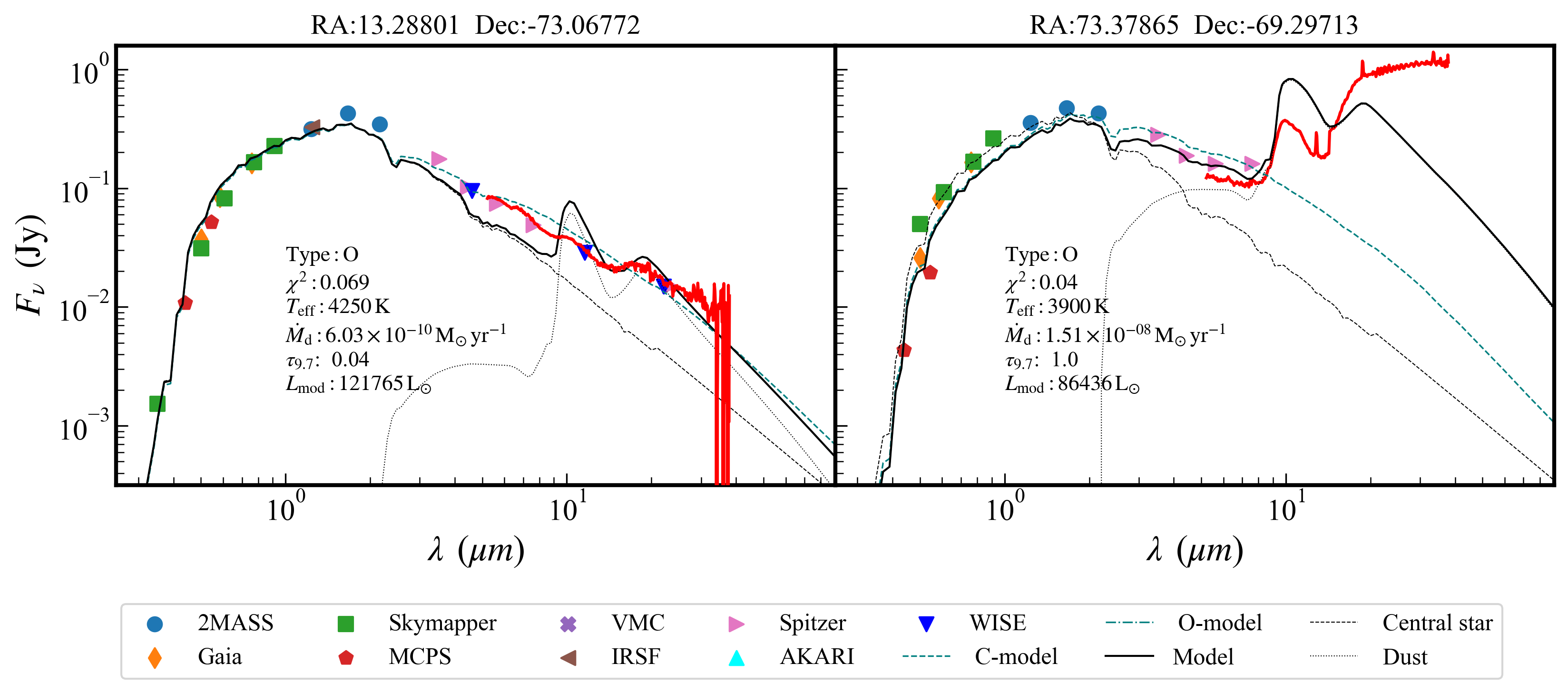}
    \caption{Two examples of modifying classification after visual inspection are shown. Observational data are denoted using various symbols and colors, while the best-fit model is represented as a solid black curve. The contributions from the central star and circumstellar dust are depicted by the black dashed and dot-dashed lines, respectively. Additionally, the blue dashed line represents the model with the minimum $\chi^2$ for another chemical type. In the left panel, the original best-fit model is C-rich (blue dashed line), which may be due to the continuous rise in the infrared band. After cross-matching with the IRS, we found that this RSG has weak but prominent silicate emission peaks at both 9.7 and 18 $\mu$m, so we changed the best-fit model (black solid line). The right panel shows a source that we moved from S1 to S2. This source has obvious infrared emission, but lacks long wavelength data, making it difficult to determine its type. Although its original best-fit model was C-rich (blue dashed line), after visual inspection, we found an rising trend at [8.0] band, which resembles an O-rich SED shape more closely. Therefore, we adjusted its classification.}
    \label{fig:6862}
\end{figure*}

We classified RSGs and AGBs into three distinct types mainly based on the model with the minimum $\chi^{2}$: O-rich (exhibiting obvious emission peaks at 9.7 and 18 $\mu m$), C-rich (displaying continuous emission in the infrared band), and optically-thin (Op-thin, showing minimal infrared emission, closely resembling the stellar atmosphere model， $\tau_{\rm 9.7/11.3} = 0.0001$). Additionally, apart from these three chemical types, we classified AGBs with 2MASS $(J-K_{\rm s})_{0} \geq 2.0$ as extreme-AGBs (x-AGBs) \citep{Boyer_20152}. Our classification approach was not solely based on the initial fitting but also included several adjustments to improve accuracy. The detailed process is as follows:  (1) In our fitting process, when dust emission was not strong enough or data beyond the [8.0] and $W3$ bands were lacking, distinguishing the chemical type based on the minimal $\chi^{2}$ became challenging. Therefore, after getting the initial classification, we re-assigned C-rich RSGs with $\tau_{\rm 11.3} \leq 0.005$ and C-rich AGBs with $\tau_{\rm 11.3} \leq 0.01$ as O-rich, because AGBs are originally O-rich and the majority of RSGs are O-rich (e.g. Paper \uppercase\expandafter{\romannumeral1}, \citealt{2023A&A...676A..84Y,2021ApJ...912..112W}). (2) To help minimize the misclassification of previous step, we partly employed the method of \citet{2018A&A...616L..13L,2019A&A...631A..24L}, which utilized Gaia and 2MASS photometric data to infer the chemical type of AGBs. Specifically，for stars were re-assigned as O-rich in previous step and their corresponding $\tau_{\rm{9.7}} \leq 0.1$, if they had $W_{RP} - W_{K} > 0.7 + 0.15 \times (K_{\rm S} - 10.8)^{2}$, we then reclassified them as C-rich. (3) For RSGs and AGBs with $\tau_{\rm {9.7/11.3}}=0.0001$, the minimum optical depth in our models, we classified them as Op-thin, regardless of whether they were C-rich or O-rich. (4) Finally, we visually inspected all fitting results and determined the types for sources with ambiguous classifications based on our experience and the Spitzer IRS Enhanced Products \citep{2004ApJS..154...18H,2015MNRAS.451.3504R}. 
This process aims to distinguish  the sources shown in Figure~\ref{fig:6862} as much as possible, which show the uncertainty and challenges in classification based on SEDs.  We will show the impact of each step of our process on the results and discuss it in Section~\ref{sec:result}.

\section{Result and Discussion}
\label{sec:result}
We present a catalog of evolved stars in the MCs, including AGBs and RSGs in the MCs, providing stellar properties such as luminosity (in this work, the luminosity is derived by integrating the observed SED ($L_{\rm{obs}}$) and integrating the best-fit model SED ($L_{\rm{mod}}$)), effective temperature, $\log{g}$, Mass, DPR and other parameters for dust shell such as chemical type, optical depth and so on. The complete table is available in machine-readable format, where  the structure and content of MRT are provided in Table~\ref{tab:result1} for reference. In this catalog, the results of RSGs in the LMC based on new models, data, and classification methods (partially referenced from \citealt{2018A&A...616L..13L}) have also been updated, with minimal changes compared to Paper \uppercase\expandafter{\romannumeral1}. In order to maintain consistency, we will use the updated results of RSGs in the LMC in the following results and discussion. Table~\ref{tab:LMCSMC} displays the statistical results of the DPR for evolved stars in the MCs. In the subsequent sections, we organize the results and discussions into two parts, focusing on RSGs and AGBs, respectively.

\begin{table}[]
  \centering
    \scriptsize
     \caption{Content of the evolved stars catalog in the MCs}
    \begin{tabular}{ll}
    \toprule[2pt]
        Column Name	& Description\\
        \midrule[2pt]
        Type & AGB or RSG, the original catalog classification from \citet{2021ApJ...923..232R} \\ 
        GALAXY & LMC or SMC  \\ 
        2MASSID & {ID from 2MASS } \\ 
        RA & Right ascension in decimal degrees (J2000)  \\ 
        DEC & Declination in decimal degrees (J2000)  \\ 
        E(B-V) & The $E(B-V)$ value come from \citet{2022MNRAS.511.1317C}  \\ 
        Jmag...Ksmag&  Initial near-IR photometric data and errors  \\ 
        fluxsku...fluxs24 & Flux after extinction correction in all bands  \\ 
        sflag & $sflag=0$ represents normal source; $sflag=-1$ and 1 represents removed source;\\ 
         & $sflag=2$ represents binary or other contamination; $sflag=3$ represents sources with bad $W3/W4$ data \\
        Teff & Best-fitting effective temperature  \\ 
        logg &  Log of surface gravity of best-fitting atmosphere model  \\ 
        Fe/H & Metallicity of best-fitting atmosphere model  \\ 
        Lummod & Log of integrated luminosity obtained from observed SED  \\ 
        Lumobs & Log of integrated luminosity obtained from best-fitting model SED  \\ 
        Mass & Mass of best-fitting atmosphere model  \\ 
        dustMass & Total mass of dust shell  \\ 
        DPR &  Best-fitting dust production rate  \\ 
        DPRC & DPR of C-rich model with minimum chi-square  \\ 
        DPRO & DPR of O-rich model with minimum chi-square\\  
        mType	&Chemical type of best-fitting model\\
        dustT	&  Temperature of dust at inner radius\\
        Rhoin	&  Density of dust at inner radius\\
        Rmin&	 Inner radius of dust shell in units of stellar radius\\
        tau&	Best-fitting optical depth\\
        tauO&	 Best-fitting O-rich model optical depth at 9.7\,$\mu$m\\
        tauC&	 Best-fitting C-rich model optical depth at 11.3\,$\mu$m\\
        PC	&The proportion of C-rich models among the 1000 models with the lowest chi-square values\\
        PO&	The proportion of O-rich models among the 1000 models with the lowest chi-square values\\
        chi-square	&Lowest chi-squared value for best-fitting model\\
        \bottomrule[2pt]
    \end{tabular}
    \label{tab:result1}
\end{table}

\begin{deluxetable*}{c|cccccc|cccccc}

\tablecaption{\label{tab:LMCSMC} Dust-production rate ($\dot{M}_{\rm d}$) of evolved stars in the MCs}
\tablehead{\multicolumn{1}{c}{}&\multicolumn{6}{c}{SMC} & \multicolumn{6}{c}{LMC}\\
\cmidrule(lr){2-7}\cmidrule(lr){8-13}
\colhead{Type}&\colhead{N}&\colhead{Mean}&\colhead{Median}&\colhead{Sum} &   \colhead{$\dot{M}_{\rm d}$\%$^a$} & \colhead{$N\%$$^{b}$} &\colhead{N}&\colhead{Mean}&\colhead{Median}&\colhead{Sum} &   \colhead{$\dot{M}_{\rm d}$\%} & \colhead{$N\%$}\\
\colhead{ }& \colhead{}  & \colhead{ ($ \rm{M_{\sun}\,{yr^{-1}}}$)}& \colhead{($\rm{ M_{\sun}\,yr^{-1}}$)} & \colhead{($\rm{ M_{\sun}\,yr^{-1}}$)}&\colhead{  }& \colhead{}& \colhead{}  & \colhead{ ($ \rm{M_{\sun}\,{yr^{-1}}}$)}& \colhead{($\rm{ M_{\sun}\,yr^{-1}}$)} & \colhead{($\rm{ M_{\sun}\,yr^{-1}}$)}&\colhead{  }&\colhead{  }}
\startdata
        AGB & 5768 & $2.84 \times10^{-10}$ & $5.08 \times10^{-11}$ & $1.64 \times10^{-06}$ & 93.84\% & 73.71\% & 30937 & $2.82 \times10^{-10}$ & $6.38 \times10^{-11}$ & $8.71 \times10^{-06}$ & 89.91\% & 86.78\%  \\ 
        RSG & 2057 & $5.23 \times10^{-11}$ & $1.2 \times10^{-12}$ & $1.08 \times10^{-07}$ & 6.16\% & 26.29\% & 4714 & $2.08 \times10^{-10}$ & $1.56 \times10^{-11}$ & $9.78 \times10^{-07}$ & 10.09\% & 13.22\%  \\ 
        Total & 7825 & $2.23 \times10^{-10}$ & $3.59 \times10^{-11}$ & $1.75 \times10^{-06}$ & 100.00\% & 100.00\% & 35651 & $2.72 \times10^{-10}$ & $5.56 \times10^{-11}$ & $9.69 \times10^{-06}$ & 100.00\% & 100.00\%  \\ 
        Total ($sflag=0$) & 7327 & $2.1 \times10^{-10}$ & $2.82 \times10^{-11}$ & $1.54 \times10^{-06}$ & 87.92\% & 93.64\% & 33298 & $2.64 \times10^{-10}$ & $5.0 \times10^{-11}$ & $8.81 \times10^{-06}$ & 90.85\% & 93.40\%  \\
\enddata
\tablecomments{$^{\rm a}$ Proportion to total DPR. $^{\rm b}$ Proportion to total number of sources. $^{\rm c} sflag=0$ represents sources that are considered normal.}
\end{deluxetable*}

\subsection{Different types of the dust shells}
\label{Different types of the dust shells}
\begin{figure*}
    \centering
	\includegraphics[width=0.9\linewidth]{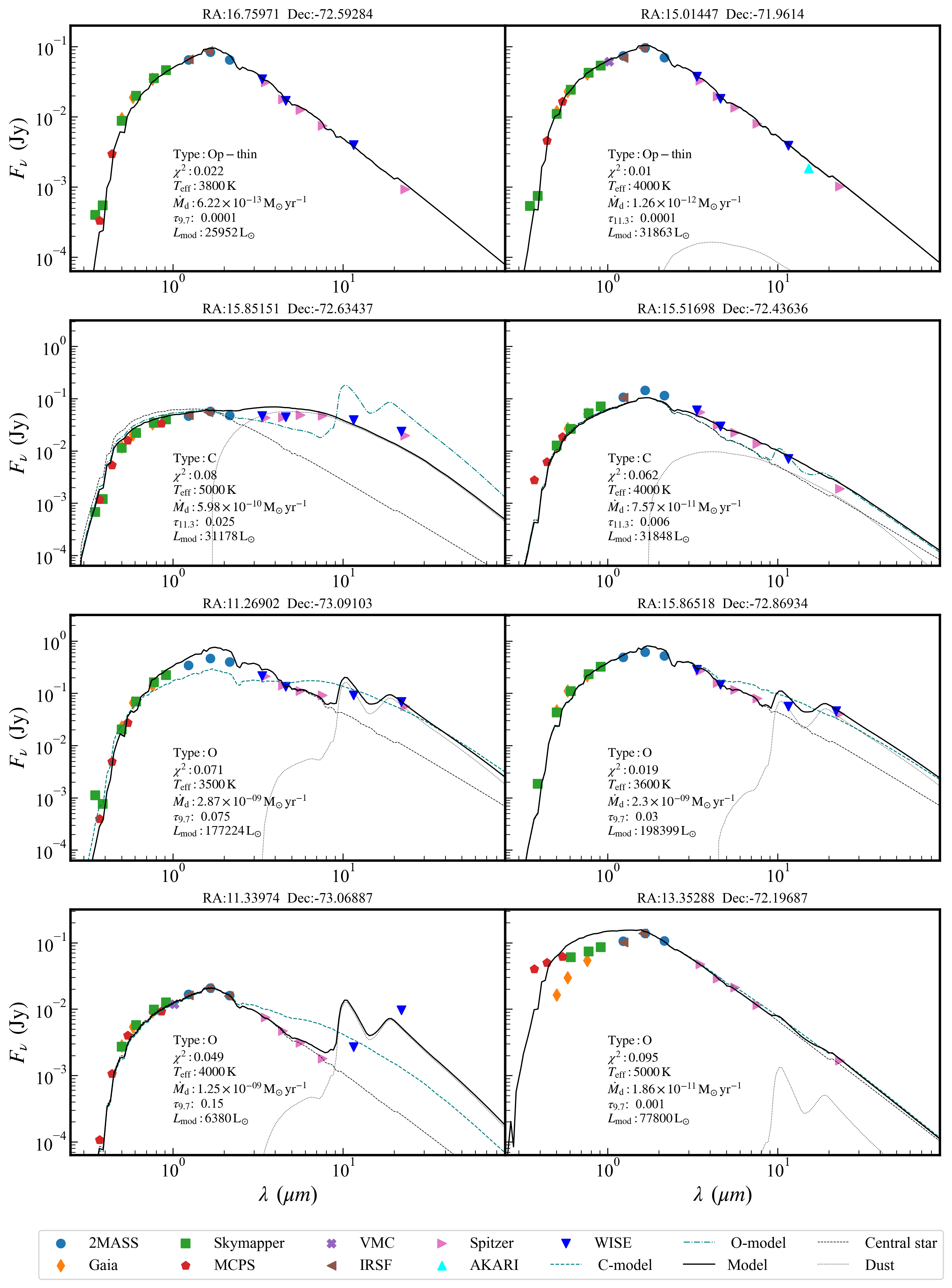}
    \caption{Examples of SED fitting for} eight RSGs in the SMC. Rows 1 to 3 show good fitting for Op-thin, C-rich, and O-rich stars ($sflag=0$), while the bottom row shows anomalous fitting due to bad $W3/W4$ data (left, $sflag=3$) and possible binary or other factors (right, $sflag=2$). 

    \label{fig:fitrsg}
\end{figure*}

\begin{figure*}
    \centering
	\includegraphics[width=0.9\linewidth]{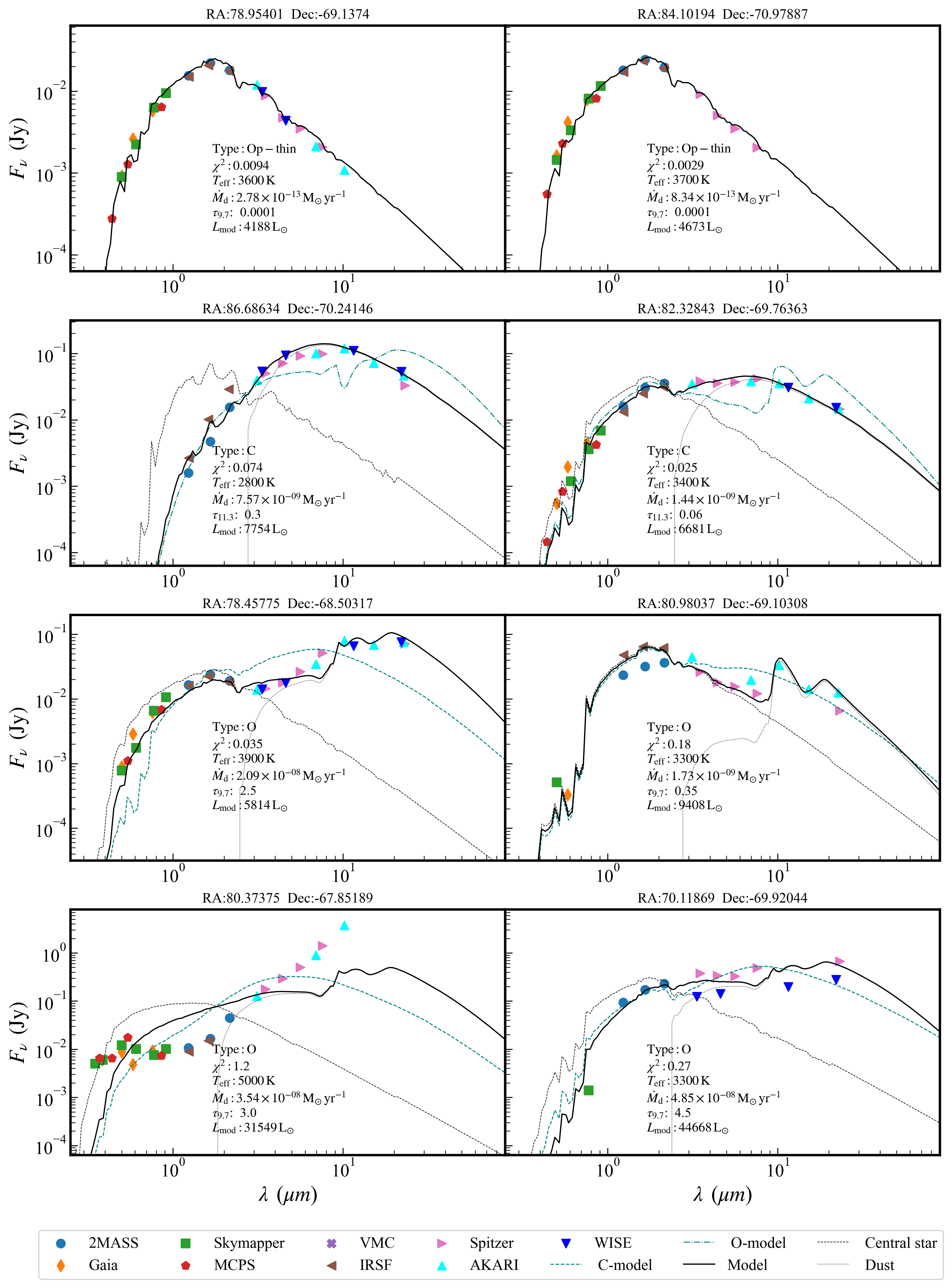}
    \caption{Examples of SED fitting for eight AGBs} in the LMC. The first, second, and third rows show good SED fitting for Op-thin, C-rich, and O-rich stars, respectively ($sflag=0$). The bottom row shows anomalous SEDs of two sources. The left panel shows a source that was removed during the data filtering process ($sflag=1$), while the right panel shows a source with possible variability ($sflag=0$).
    \label{fig:fitagb}
\end{figure*}

We established a ``$sflag$" classification, where normal sources were assigned $sflag=0$. The sources removed in Section~\ref{sec:data} during data filtering are marked as $sflag=1$ or $-1$. Sources with $sflag=2$ fall into two categories: 1) those we consider may not be evolved stars, showing distinctly abnormal SEDs, despite not being identified as other contaminants in Simbad; 2) those with poorly fitted SEDs, possibly influenced by binaries or variability. $sflag=3$ denotes sources with bad data in the $W3$ and $W4$ bands. Figure~\ref{fig:fitrsg} and Figure~\ref{fig:fitagb} show several examples of RSGs and AGBs fitting result, with different rows representing different types, the best-fit model is represented as a solid black curve. The first to third rows represent Op-thin, C-rich, and O-rich ($sflag=0$), respectively. The bottom row of Figure~\ref{fig:fitrsg} display the SEDs of two examples of poorly fitted RSGs, possibly resulting from abnormal $W3/W4$ data (left, $sflag=3$) or binary and other factors (right, $sflag=2$). The left panel in the bottom row of Figure~\ref{fig:fitagb} shows a removed source ($sflag=1$), while the right panel depicts an AGBs with possible variability ($sflag=0$).

\begin{figure*}
	\centerline{\includegraphics[width=0.9\linewidth]{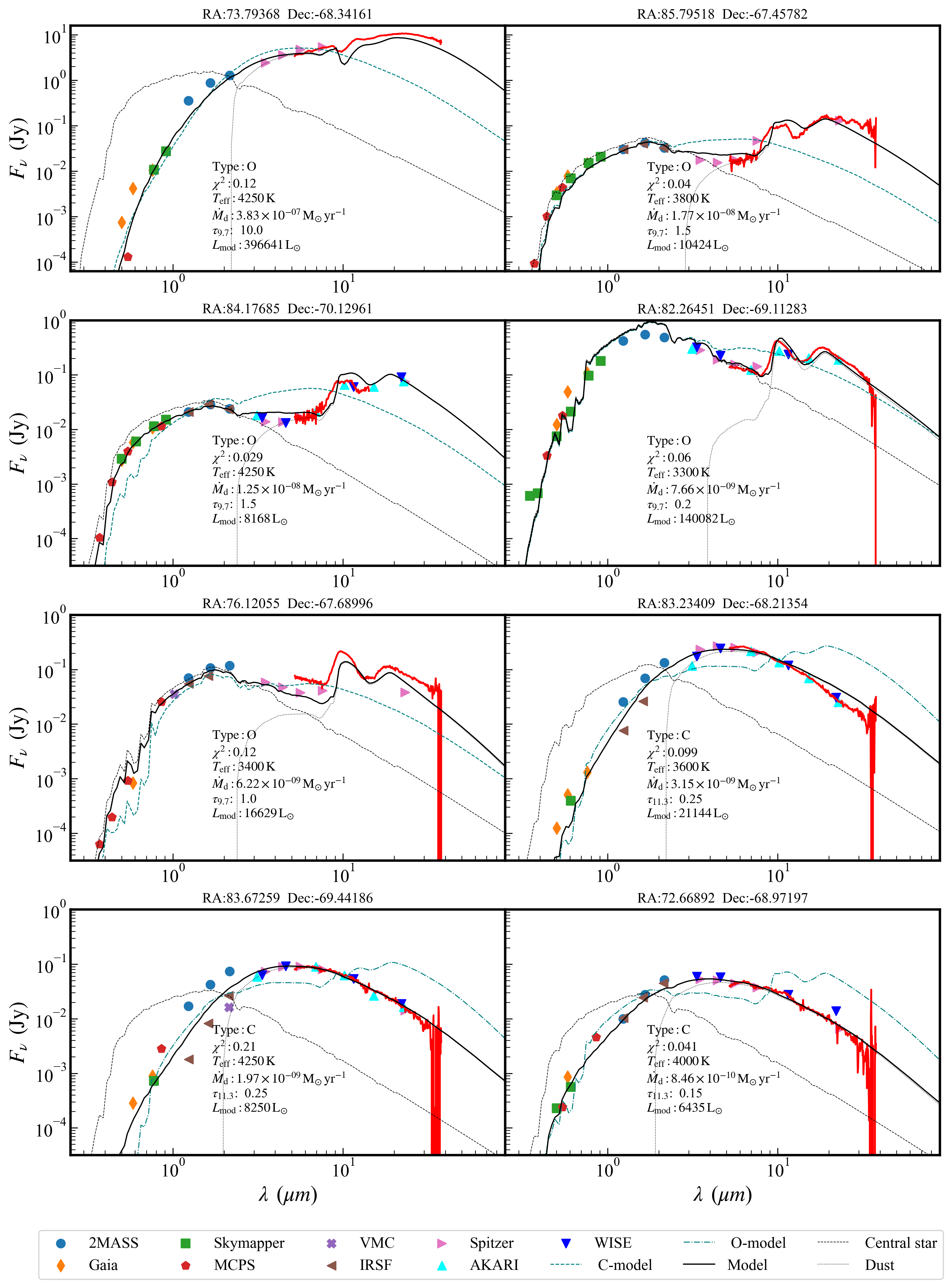}}
    \caption{Examples of high DPR stars in our sample. The red line represents the IRS spectral data.}
    \label{fig:specagb}
\end{figure*}

To verify our classification results, we performed a cross-match with Spitzer IRS (InfraRed Spectrograph) Enhanced Products \citep{2004ApJS..154...18H,2015MNRAS.451.3504R}, using a cross-matching radius of 3$^{\prime\prime}$. A total of 203 sources in the LMC and 76 sources in the SMC were matched with IRS Enhanced Products, and 32 (LMC) and 19 (SMC) of them were misclassified, which indicated that in most cases, our fitting results matched well with the spectra, as shown in Figure~\ref{fig:specagb}. The misclassification can be attributed to the ambiguity in the dust features, as shown in the left panel of Figure~\ref{fig:6862}.  We also compared our classification results with those from GAIA DR3 \citep[i.e.][]{2023A&A...674A...1G,2023A&A...674A..39G,2023A&A...674A..15L}. We cross-matched our catalog with the``gold\_sample\_carbon\_stars" catalog of GAIA DR3, for which there are 5,293 and 938 common sources in the LMC and SMC, with corresponding ``C-rich'' stars of 96.39\% and 87.42\% from our work, respectively. For the ``CSTAR" label from Gaia DR3, the numbers are 5,303 (96.30\%) for the LMC and 940 (87.34\%) for the SMC, respectively.
It's worth noting that distinguishing the chemical types of sources with relatively small optical depth based solely on photometric data is challenging. Therefore, our chemical classification is a actually rough  estimation.

\subsubsection{Classification of RSGs}
The chemical classification of RSGs in the LMC and SMC are presented in Table~\ref{tab:DPRlmcrsg} and Table~\ref{tab:DPRsmcrsg}, respectively. The SMC has a higher proportion ($\sim$53\%) of Op-thin RSGs than the LMC ($\sim$39\%), which indicates that as metallicity decreases, the fraction of naked stars increases \citep{2008ApJ...686.1056S}. 

\begin{deluxetable*}{cccccccccc}
\tabletypesize{\footnotesize}
\tablecaption{\label{tab:DPRlmcrsg} Dust-production rate ($\dot{M}_{\rm d}$) of RSGs in the LMC}
\tablehead{\colhead{Type$^a$} & \colhead{N} & \colhead{Mean} &\colhead{ Median} & \colhead{Max}  & \colhead{Min}  & \colhead{Sum} &  \colhead{$\dot{M}_{\rm d}$\%$^b$} &  \colhead{$N\%$$^{c}$}\\
\colhead{ }& \colhead{}  & \colhead{ ($ \rm{M_{\sun}\,{yr^{-1}}}$)}& \colhead{($\rm{ M_{\sun}\,yr^{-1}}$)} & \colhead{($\rm{ M_{\sun}\,yr^{-1}}$)}&\colhead{ ($\rm{ M_{\sun}\,yr^{-1}}$) }& \colhead{($\rm{ M_{\sun}\,yr^{-1}}$)}&\colhead{ } & \colhead{ } }
\startdata
        C & 350 & $8.43 \times10^{-11}$ & $4.49 \times10^{-11}$ & $1.31 \times10^{-09}$ & $1.56 \times10^{-12}$ & $2.95 \times10^{-08}$ & 3.01\% & 7.42\%  \\ 
        $sflag=0 $ & 338 & $8.58 \times10^{-11}$ & $4.53 \times10^{-11}$ & $1.31 \times10^{-09}$ & $1.56 \times10^{-12}$ & $2.9 \times10^{-08}$ & 3.23\% & 7.58\%  \\ \hline
        O & 2508 & $3.78 \times10^{-10}$ & $3.89 \times10^{-11}$ & $2.26 \times10^{-08}$ & $7.71 \times10^{-13}$ & $9.48 \times10^{-07}$ & 96.86\% & 53.20\%  \\ 
        $sflag=0$ & 2274 & $3.81 \times10^{-10}$ & $3.48 \times10^{-11}$ & $2.26 \times10^{-08}$ & $7.71 \times10^{-13}$ & $8.67 \times10^{-07}$ & 96.63\% & 50.96\%  \\ \hline
        Op-thin & 1856 & $6.53 \times10^{-13}$ & $5.9 \times10^{-13}$ & $1.09 \times10^{-11}$ & $1.97 \times10^{-13}$ & $1.21 \times10^{-09}$ & 0.12\% & 39.37\%  \\ 
        $sflag=0$ & 1850 & $6.53 \times10^{-13}$ & $5.9 \times10^{-13}$ & $1.09 \times10^{-11}$ & $1.97 \times10^{-13}$ & $1.21 \times10^{-09}$ & 0.13\% & 41.46\%  \\ \hline
        All & 4714 & $2.08 \times10^{-10}$ & $1.56 \times10^{-11}$ & $2.26 \times10^{-08}$ & $1.97 \times10^{-13}$ & $9.78 \times10^{-07}$ & 100.00\% & 100.00\%  \\ 
        $sflag=0$ & 4462 & $2.01 \times10^{-10}$ & $1.1 \times10^{-11}$ & $2.26 \times10^{-08}$ & $1.97 \times10^{-13}$ & $8.97 \times10^{-07}$ & 91.66\% & 94.65\%  \\ 
\enddata
\tablecomments{$^{\rm a}$ $sflag=0$ represents sources that are considered normal. For each type, upper row indicates the numbers related to all RSGs, while bottom row indicates the numbers related to RSGs with $sflag=0$.  $^{\rm b}$ Proportion to total DPR. $^{\rm c}$ Proportion to total number of RSGs. The proportions of the last row  is ratio of RSGs with $sflag =0$ to all RSGs.} 
\end{deluxetable*}

\begin{deluxetable*}{cccccccccc}
\tabletypesize{\footnotesize}
\tablecaption{\label{tab:DPRsmcrsg} Dust-production rate ($\dot{M}_{\rm d}$) of RSGs in the SMC}
\tablehead{\colhead{Type$^a$} & \colhead{N} & \colhead{Mean} &\colhead{ Median} & \colhead{Max}  & \colhead{Min}  & \colhead{Sum} &  \colhead{$\dot{M}_{\rm d}$\%$^b$} &  \colhead{$N\%$$^{c}$}\\
\colhead{ }& \colhead{}  & \colhead{ ($ \rm{M_{\sun}\,{yr^{-1}}}$)}& \colhead{($\rm{ M_{\sun}\,yr^{-1}}$)} & \colhead{($\rm{ M_{\sun}\,yr^{-1}}$)}&\colhead{ ($\rm{ M_{\sun}\,yr^{-1}}$) }& \colhead{($\rm{ M_{\sun}\,yr^{-1}}$)}&\colhead{ } & \colhead{ } }
\startdata
        C & 89 & $9.54 \times10^{-11}$ & $5.53 \times10^{-11}$ & $8.74 \times10^{-10}$ & $2.21 \times10^{-12}$ & $8.49 \times10^{-09}$ & 7.90\% & 4.33\%  \\ 
        $sflag=0$ & 84 & $9.84 \times10^{-11}$ & $5.53 \times10^{-11}$ & $8.74 \times10^{-10}$ & $2.21 \times10^{-12}$ & $8.26 \times10^{-09}$ & 9.84\% & 4.30\%  \\ \hline
        O & 882 & $1.11 \times10^{-10}$ & $3.85 \times10^{-11}$ & $4.02 \times10^{-09}$ & $1.28 \times10^{-12}$ & $9.83 \times10^{-08}$ & 91.37\% & 42.88\%  \\ 
        $sflag=0$ & 785 & $9.55 \times10^{-11}$ & $3.3 \times10^{-11}$ & $4.02 \times10^{-09}$ & $1.28 \times10^{-12}$ & $7.49 \times10^{-08}$ & 89.22\% & 40.15\%  \\ \hline
        Op-thin & 1086 & $7.29 \times10^{-13}$ & $5.9 \times10^{-13}$ & $5.38 \times10^{-12}$ & $2.75 \times10^{-13}$ & $7.92 \times10^{-10}$ & 0.74\% & 52.80\%  \\ 
        $sflag=0$ & 1086 & $7.29 \times10^{-13}$ & $5.9 \times10^{-13}$ & $5.38 \times10^{-12}$ & $2.75 \times10^{-13}$ & $7.92 \times10^{-10}$ & 0.94\% & 55.55\%  \\ \hline
        All & 2057 & $5.23 \times10^{-11}$ & $1.2 \times10^{-12}$ & $4.02 \times10^{-09}$ & $2.75 \times10^{-13}$ & $1.08 \times10^{-07}$ & 100.00\% & 100.00\%  \\ 
        $sflag=0$ & 1955 & $4.3 \times10^{-11}$ & $1.08 \times10^{-12}$ & $4.02 \times10^{-09}$ & $2.75 \times10^{-13}$ & $8.4 \times10^{-08}$ & 78.10\% & 95.04\%  \\
\enddata
\tablecomments{$^{\rm a}$ $sflag=0$ represents sources that are considered normal. For each type, upper row indicates the numbers related to all RSGs, while bottom row indicates the numbers related to RSGs with $sflag=0$.  $^{\rm b}$ Proportion to total DPR. $^{\rm c}$ Proportion to total number of RSGs. The proportions of the last row  is ratio of RSGs with $sflag =0$ to all RSGs.}
\end{deluxetable*}

\begin{figure*}
	\centerline{\includegraphics[width=0.9\linewidth]{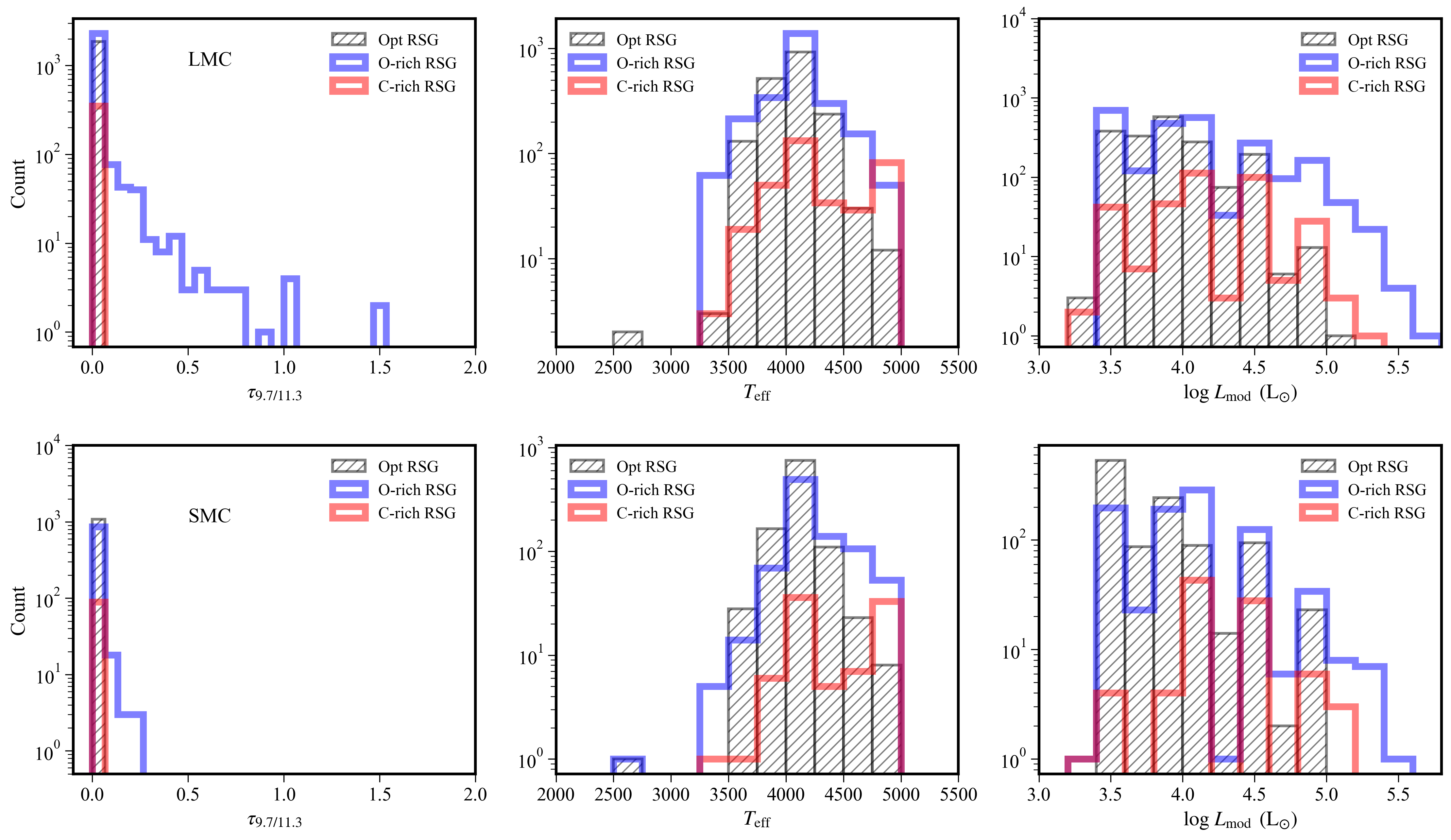}}
    \caption{Histograms of the distribution of optical depth (left), effective temperature (center), and luminosity (right) for RSGs in the MCs. All RSGs with $sflag\neq 1$ or $-1$ are included. The gray line represents opt-thin RSGs, the blue line represents O-rich RSGs, the red line represents C-rich RSGs.}
    \label{fig:RSGtefflumtau}
\end{figure*}

Figures~\ref{fig:RSGtefflumtau} depicts histograms of optical depth, effective temperature, and luminosity for  RSGs in the MCs. Despite employing over 70,000 templates, which include low-temperature and low-luminosity AGBs models for fitting, over 90\% of RSGs in the MCs exhibit effective temperatures between 3000\,K and 4500\,K, and luminosities ranging from 3,000$\,\rm L_{\sun}$ to 300,000$\,\rm L_{\sun}$. There are some clear differences between the LMC and SMC. The SMC lacks RSGs with higher $\tau_{\rm{9.7}}$, indicating that the formation of O-rich dust is more challenging at low metallicity \citep{2000A&A...354..125V,2006ASPC..353..211V,2008ApJ...686.1056S}. Compared with the LMC, the overall effective temperature of RSGs in the SMC is slightly higher, with average values of 4075\,K for the SMC and 4009\,K for the LMC. \citet{2003AJ....126.2867M,2006ApJ...645.1102L,2005ApJ...628..973L} found that the distribution of spectral types is skewed toward earlier type at lower metallicitiy and suggest that the effect of metallicity on the appearance of TiO lines is likely to account for this difference.

\subsubsection{Classification of AGBs}
\begin{figure*}
	\centerline{\includegraphics[width=0.9\linewidth]{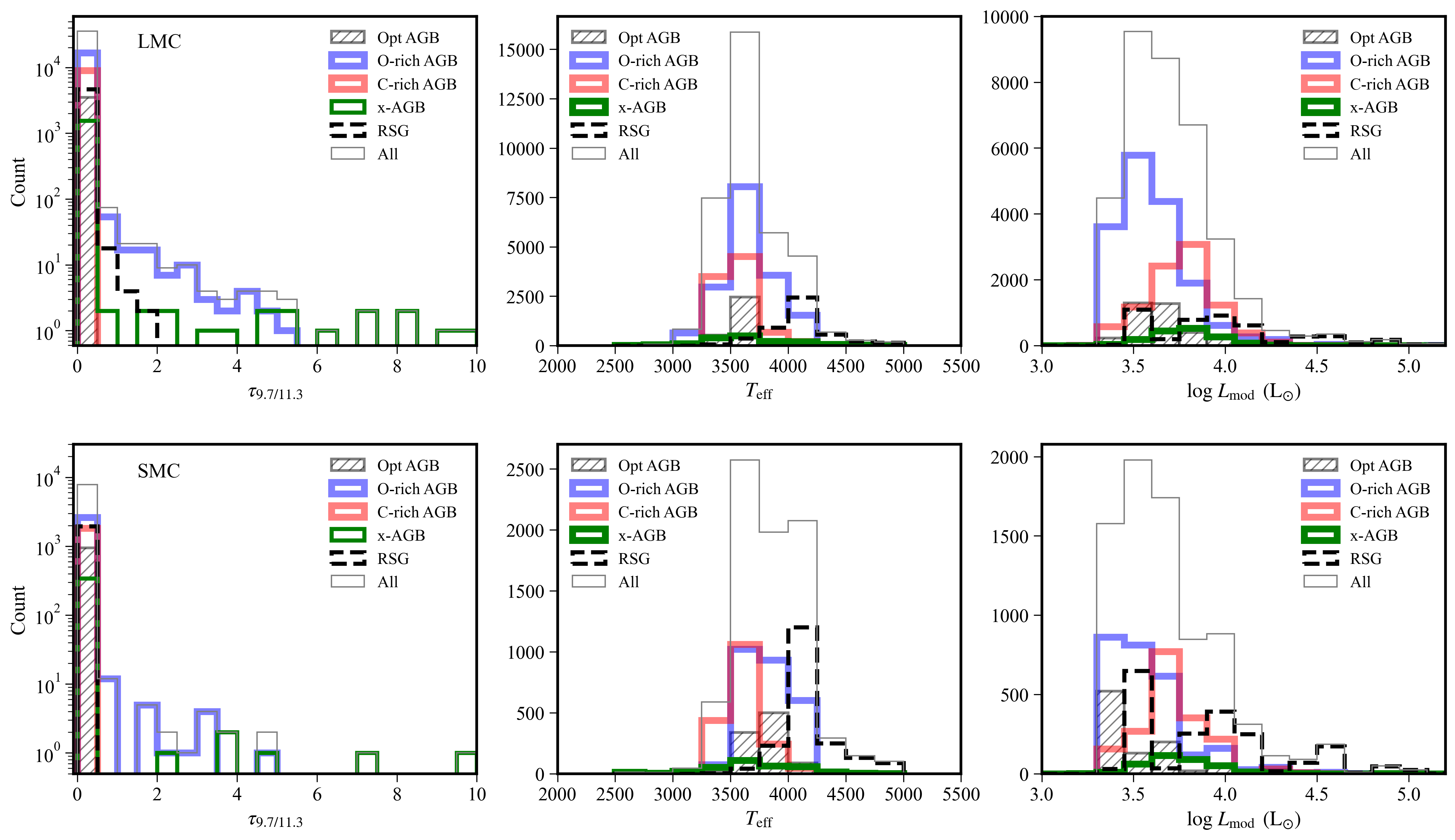}}
    \caption{Histograms of the distribution of optical depth (left), effective temperature (center), and luminosity (right) for AGBs and RSGs in the MCs. All sources with $sflag\neq 1$ or $-1$ are included. The gray filled line represents optically thin AGBs, the blue line represents O-rich AGBs, the red line represents C-rich AGBs, the green line represents x-AGBs, the black dashed-dotted line represents all RSGs (regardless of chemical type), and the gray line represents all AGBs and RSGs.}
    \label{fig:tefflumtau}
\end{figure*}

\begin{figure*}
	\centerline{\includegraphics[width=0.8\linewidth]{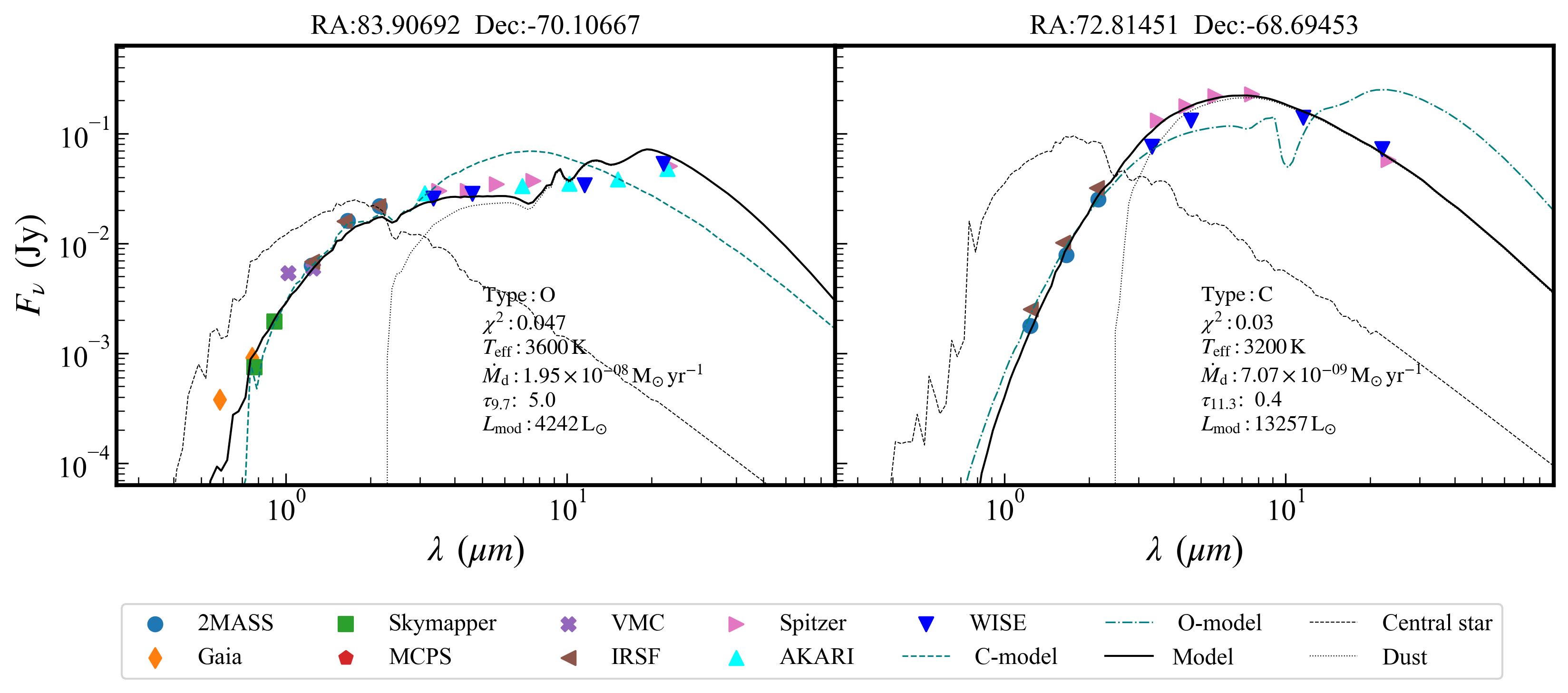}}
    \caption{Two typical examples of SED fitting for x-AGBs, with x-O on the left and x-C  on the right.}
    \label{fig:xagb}
\end{figure*}

The chemical classification results for AGBs in the LMC and SMC are presented in Table~\ref{tab:DPRlmcagb} and Table~\ref{tab:DPRsmcagb}，  respectively. We designated sources with 2MASS $(J-K_{\rm s})_{0} \geq 2.0$ as x-AGB. In the LMC, there are 1,570 x-AGBs, of which 1,547 are C-rich (x-C, as shown in the right panel of Figure~\ref{fig:xagb}), and only 23 are O-rich (x-O, as shown in the left panel of Figure~\ref{fig:xagb}), the majority of x-AGBs are carbon stars. Only 11\% of the AGBs in the LMC were classified as Op-thin, which is lower than that of RSGs. The SMC exhibited a higher proportion of Op-thin AGBs compared to the LMC, constituting 16\% of the total number. 
As shown in Figure~\ref{fig:tefflumtau}, the overall effective temperature of AGBs in the SMC is slightly higher than that of LMC, which is consist with RSGs. The average effective temperatures of AGB in the LMC and SMC we obtained are 3590\,K and 3697\,K, respectively. Both in the LMC and SMC, C-rich AGBs seem to have higher luminosity distribution peak than O-rich AGBs. C-rich stars are generally more evolved than O-rich, having undergone several thermal pulses became C-rich and luminosity also increases with age 

In our results, the proportion of C-rich and x-C in the SMC is slightly higher than that in the LMC, but it is not very obvious. Previous studies on AGB stars have shown that the ratio of C-rich to O-rich AGBs varies significantly across different environments and is correlated with metallicity: lower metallicity leads to more C-rich AGBs \citep{1980ApJ...242..938B,1986ApJ...305..634C,2003A&A...402..133C}. \citet{1983ApJ...275L..65I} explained the relation between metallicity and the C/M ratio, noting that at lower metallicity, O-rich AGBs more readily transform into C-rich AGBs and stay in the AGB phase longer because fewer carbon atoms are required for the transformation. They also explained the relation between metallicity and stellar temperature. At lower metallicity, the evolutionary tracks of AGB stars shift to higher temperatures, the abundance of TiO molecules decreases, and the atmosphere becomes more transparent. Although we found that our results are consistent with the above studies, there is considerable uncertainty in the classification of AGBs.

\begin{deluxetable*}{cccccccccc}
\tablewidth{0pt} 
\tabletypesize{}
\tablecaption{\label{tab:DPRlmcagb} Dust-production rate ($\dot{M}_{\rm d}$) of AGBs in the LMC}
\tablehead{\colhead{Type $^{a}$} & \colhead{N} & \colhead{Mean} &\colhead{ Median} & \colhead{Max}  & \colhead{Min}  & \colhead{Sum} &  \colhead{$\dot{M}_{\rm d} $\% $^b$} &  \colhead{$N\%$ $^{c}$}\\
\colhead{ }& \colhead{}  & \colhead{ ($ \rm{M_{\sun}\,{yr^{-1}}}$)}& \colhead{($\rm{ M_{\sun}\,yr^{-1}}$)} & \colhead{($\rm{ M_{\sun}\,yr^{-1}}$)}&\colhead{ ($\rm{ M_{\sun}\,yr^{-1}}$) }& \colhead{($\rm{ M_{\sun}\,yr^{-1}}$)}&\colhead{ } & \colhead{ } }
\startdata
        C & 9027 & $1.26 \times10^{-10}$ & $1.02 \times10^{-10}$ & $2.27 \times10^{-09}$ & $1.53 \times10^{-12}$ & $1.14 \times10^{-06}$ & 13.07\% & 29.18\%  \\ 
        $sflag=0$ & 8848 & $1.27 \times10^{-10}$ & $1.02 \times10^{-10}$ & $2.27 \times10^{-09}$ & $1.53 \times10^{-12}$ & $1.13 \times10^{-06}$ & 14.25\% & 30.68\%  \\ \hline
        O & 16840 & $1.8 \times10^{-10}$ & $5.0 \times10^{-11}$ & $6.7 \times10^{-08}$ & $5.5 \times10^{-13}$ & $3.04 \times10^{-06}$ & 34.83\% & 54.43\%  \\ 
        $sflag=0$ & 14936 & $1.53 \times10^{-10}$ & $4.17 \times10^{-11}$ & $6.7 \times10^{-08}$ & $5.5 \times10^{-13}$ & $2.29 \times10^{-06}$ & 28.90\% & 51.80\%  \\ \hline
        Op-thin & 3500 & $5.57 \times10^{-13}$ & $3.93 \times10^{-13}$ & $3.78 \times10^{-12}$ & $1.97 \times10^{-13}$ & $1.95 \times10^{-09}$ & 0.02\% & 11.31\%  \\ 
        $sflag=0$ & 3492 & $5.56 \times10^{-13}$ & $3.93 \times10^{-13}$ & $3.78 \times10^{-12}$ & $1.97 \times10^{-13}$ & $1.94 \times10^{-09}$ & 0.02\% & 12.11\%  \\ \hline
        x-C & 1547 & $1.66 \times10^{-09}$ & $8.46 \times10^{-10}$ & $1.7 \times10^{-08}$ & $1.02 \times10^{-10}$ & $2.57 \times10^{-06}$ & 29.46\% & 5.00\%  \\ 
        $sflag=0$ & 1540 & $1.66 \times10^{-09}$ & $8.46 \times10^{-10}$ & $1.7 \times10^{-08}$ & $1.02 \times10^{-10}$ & $2.56 \times10^{-06}$ & 32.34\% & 5.34\%  \\ \hline
        x-O & 23 & $8.57 \times10^{-08}$ & $4.97 \times10^{-08}$ & $5.17 \times10^{-07}$ & $1.53 \times10^{-09}$ & $1.97 \times10^{-06}$ & 22.62\% & 0.07\%  \\
        $sflag=0$ & 20 & $9.68 \times10^{-08}$ & $5.41 \times10^{-08}$ & $5.17 \times10^{-07}$ & $4.17 \times10^{-09}$ & $1.94 \times10^{-06}$ & 24.48\% & 0.07\%  \\ \hline
        x-AGB & 1570 & $2.89 \times10^{-09}$ & $8.46 \times10^{-10}$ & $5.17 \times10^{-07}$ & $1.02 \times10^{-10}$ & $4.54 \times10^{-06}$ & 52.08\% & 5.07\%  \\
        $sflag=0$ & 1560 & $2.88 \times10^{-09}$ & $8.46 \times10^{-10}$ & $5.17 \times10^{-07}$ & $1.02 \times10^{-10}$ & $4.49 \times10^{-06}$ & 56.82\% & 5.41\%  \\ \hline
        All & 30937 & $2.82 \times10^{-10}$ & $6.38 \times10^{-11}$ & $5.17 \times10^{-07}$ & $1.97 \times10^{-13}$ & $8.71 \times10^{-06}$ & 100.00\% & 100.00\%  \\ 
        All($sflag=0$) & 28836 & $2.74 \times10^{-10}$ & $5.56 \times10^{-11}$ & $5.17 \times10^{-07}$ & $1.97 \times10^{-13}$ & $7.91 \times10^{-06}$ & 90.76\% & 93.21\%  \\
\enddata
\tablecomments{$^{\rm a}$ $sflag=0$ represents sources that are considered normal. For each type, upper row indicates the numbers related to all AGBs, while bottom row indicates the numbers related to AGBs with $sflag=0$.  $^{\rm b}$ Proportion to total DPR. $^{\rm c}$ Proportion to total number of AGBs. The proportions of the last row  is ratio of AGBs with $sflag =0$ to all AGBs.}
\end{deluxetable*}

\begin{deluxetable*}{cccccccccc}
\tabletypesize{\footnotesize}
\tablecaption{\label{tab:DPRsmcagb} Dust-production rate ($\dot{M}_{\rm d}$) of AGBs in the SMC}
\tablehead{\colhead{Type $^{a}$} & \colhead{N} & \colhead{Mean} &\colhead{ Median} & \colhead{Max}  & \colhead{Min}  & \colhead{Sum} &  \colhead{$\dot{M}_{\rm d}$\% $^b$} &  \colhead{$N\%$ $^{c}$}\\
\colhead{ }& \colhead{}  & \colhead{ ($ \rm{M_{\sun}\,{yr^{-1}}}$)}& \colhead{($\rm{ M_{\sun}\,yr^{-1}}$)} & \colhead{($\rm{ M_{\sun}\,yr^{-1}}$)}&\colhead{ ($\rm{ M_{\sun}\,yr^{-1}}$) }& \colhead{($\rm{ M_{\sun}\,yr^{-1}}$)}&\colhead{ } & \colhead{ } }
\startdata
        C & 1820 & $1.15 \times10^{-10}$ & $9.58 \times10^{-11}$ & $2.65 \times10^{-09}$ & $1.53 \times10^{-12}$ & $2.1 \times10^{-07}$ & 12.81\% & 31.55\%  \\ 
        $sflag=0$ & 1747 & $1.18 \times10^{-10}$ & $9.58 \times10^{-11}$ & $2.65 \times10^{-09}$ & $1.53 \times10^{-12}$ & $2.06 \times10^{-07}$ & 14.18\% & 32.52\%  \\ \hline
        O & 2654 & $1.73 \times10^{-10}$ & $3.93 \times10^{-11}$ & $2.8 \times10^{-08}$ & $1.1 \times10^{-12}$ & $4.58 \times10^{-07}$ & 27.96\% & 46.01\%  \\ 
        $sflag=0$ & 2334 & $1.43 \times10^{-10}$ & $3.32 \times10^{-11}$ & $2.8 \times10^{-08}$ & $1.1 \times10^{-12}$ & $3.33 \times10^{-07}$ & 22.95\% & 43.45\%  \\ \hline
        Op-thin & 947 & $6.47 \times10^{-13}$ & $5.9 \times10^{-13}$ & $2.39 \times10^{-12}$ & $1.97 \times10^{-13}$ & $6.13 \times10^{-10}$ & 0.04\% & 16.42\%  \\ 
        $sflag=0$ & 946 & $6.47 \times10^{-13}$ & $5.9 \times10^{-13}$ & $2.39 \times10^{-12}$ & $1.97 \times10^{-13}$ & $6.12 \times10^{-10}$ & 0.04\% & 17.61\%  \\ \hline
        x-C & 341 & $1.76 \times10^{-09}$ & $1.13 \times10^{-09}$ & $1.35 \times10^{-08}$ & $6.77 \times10^{-11}$ & $6.01 \times10^{-07}$ & 36.70\% & 5.91\%  \\ 
        $sflag=0$ & 341 & $1.76 \times10^{-09}$ & $1.13 \times10^{-09}$ & $1.35 \times10^{-08}$ & $6.77 \times10^{-11}$ & $6.01 \times10^{-07}$ & 41.44\% & 6.35\%  \\ \hline
        x-O & 6 & $6.14 \times10^{-08}$ & $3.31 \times10^{-08}$ & $2.15 \times10^{-07}$ & $9.73 \times10^{-09}$ & $3.68 \times10^{-07}$ & 22.49\% & 0.10\%  \\ 
        $sflag=0$ & 4 & $7.76 \times10^{-08}$ & $4.29 \times10^{-08}$ & $2.15 \times10^{-07}$ & $9.73 \times10^{-09}$ & $3.1 \times10^{-07}$ & 21.39\% & 0.07\%  \\  \hline
        x-AGB & 347 & $2.79 \times10^{-09}$ & $1.14 \times10^{-09}$ & $2.15 \times10^{-07}$ & $6.77 \times10^{-11}$ & $9.7 \times10^{-07}$ & 59.19\% & 6.02\%  \\ 
        $sflag=0$ & 345 & $2.64 \times10^{-09}$ & $1.13 \times10^{-09}$ & $2.15 \times10^{-07}$ & $6.77 \times10^{-11}$ & $9.12 \times10^{-07}$ & 62.84\% & 6.42\%  \\ \hline
        All & 5768 & $2.84 \times10^{-10}$ & $5.08 \times10^{-11}$ & $2.15 \times10^{-07}$ & $1.97 \times10^{-13}$ & $1.64 \times10^{-06}$ & 100.00\% & 100.00\%  \\ 
        All(s=0) & 5372 & $2.7 \times10^{-10}$ & $4.91 \times10^{-11}$ & $2.15 \times10^{-07}$ & $1.97 \times10^{-13}$ & $1.45 \times10^{-06}$ & 88.57\% & 93.13\%  \\ 
\enddata
\tablecomments{$^{\rm a}$ $sflag=0$ represents sources that are considered normal. For each type, upper row indicates the numbers related to all AGBs, while bottom row indicates the numbers related to AGBs with $sflag=0$.  $^{\rm b}$ Proportion to total DPR. $^{\rm c}$ Proportion to total number of AGBs. The proportions of the last row  is ratio of AGBs with $sflag =0$ to all AGBs.}
\end{deluxetable*}

\subsection{The resultant DPR}
\subsubsection{DPR of RSGs}

\begin{figure*}
    \centering
	\includegraphics[width=0.8\linewidth]{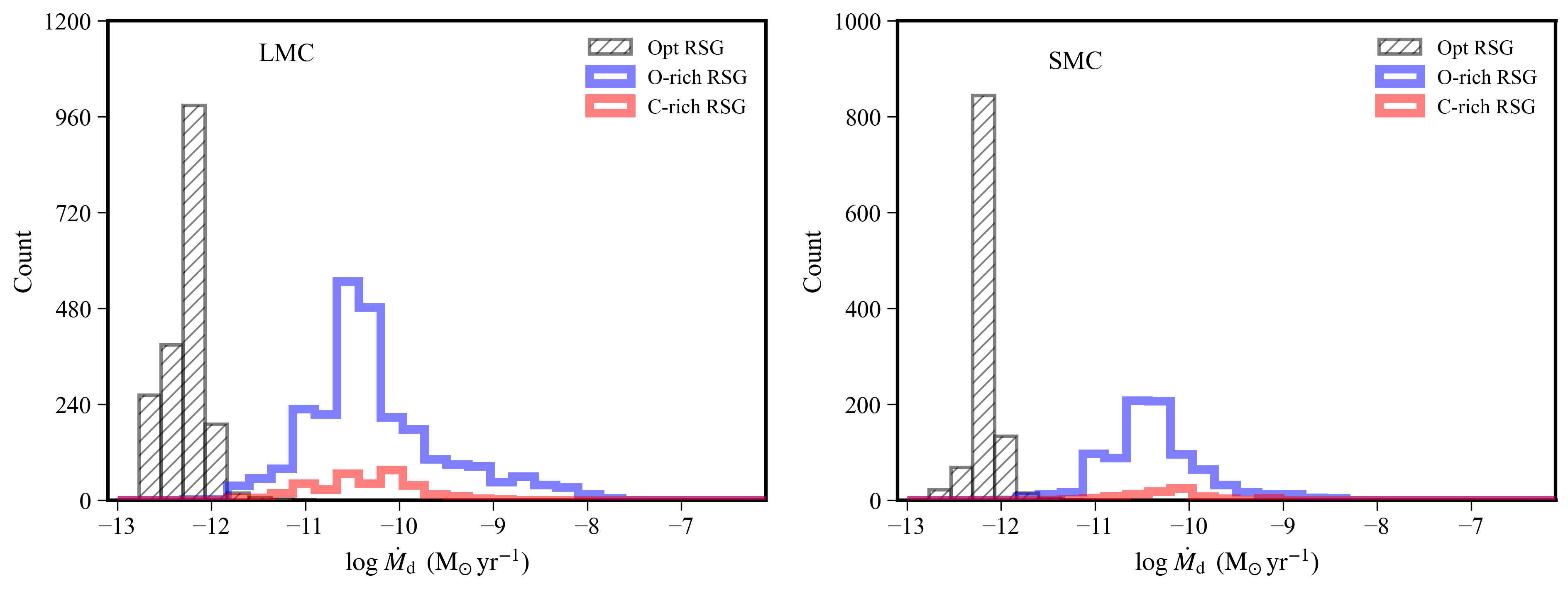}
    \caption{Histograms of the RSGs DPR ($\dot{M_{\rm d}}$) distribution in the MCs. Different colors represent different types. }
    \label{fig:rsgdpr}
\end{figure*}

The histograms of DPR distribution for RSGs  were presented in Figure~\ref{fig:rsgdpr}, detailed DPR values were provided in Table~\ref{tab:DPRlmcrsg} and \ref{tab:DPRsmcrsg}. The total DPR for RSGs in the SMC is approximately $1.08\times 10^{-7}\, \rm{M_{\odot}\,yr^{-1}}$, with an average DPR is $5.23\times 10^{-11}\, \rm{M_{\odot}\,yr^{-1}}$, lower than that in the LMC ($2.08\times 10^{-10}\, \rm{M_{\odot}\,yr^{-1}}$).  This difference may be attributed to the higher proportion of Op-thin RSGs in the SMC, which might be related to metallicity, as the metallicity decreases, the fraction of naked stars increases \citep{2008ApJ...686.1056S}, the number of high DPR O-rich stars decreases, and the mean DPR of RSGs decreases.
 
\begin{figure*}
    \centering
	\includegraphics[width=0.8\linewidth]{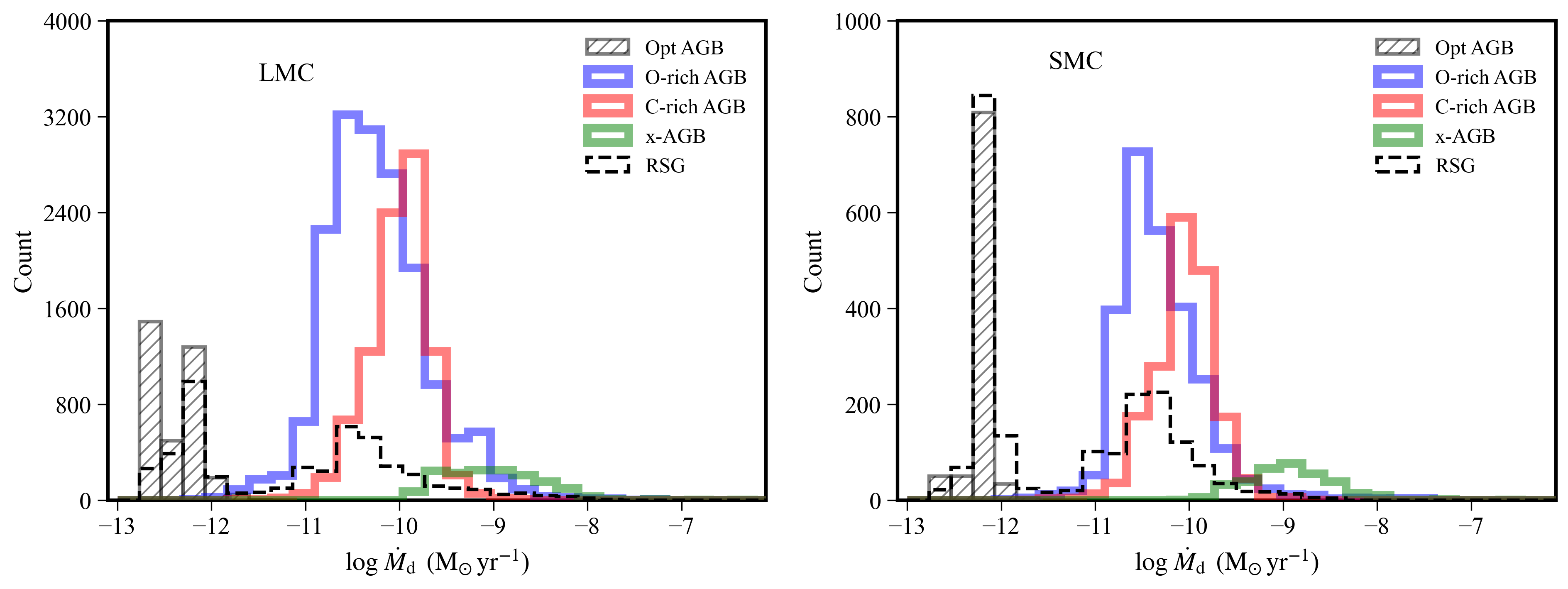}
    \caption{Histograms of the AGBs DPR ($\dot{M_{\rm d}}$) distribution in the MCs. Different colors represent different types. }
    \label{fig:agbdpr}
\end{figure*}

 Op-thin RSGs and those with low optical depth exhibit extremely low DPR value ($\dot{M}_{\rm d}\leq 10^{-11}\, \rm{M{\odot}\,yr^{-1}}$) because these RSGs are fitted to the lower limit of our template's DPR, which may not accurately reflect their actual DPR and MLR values. For example,  SED with $\tau_{\rm{9.7/11.3}}=0.001$ and $\tau_{\rm{9.7/11.3}}=0.0001$ may be almost indistinguishable, but their DPR values may differ by an order of magnitude, which is an important source of uncertainty for DPR of sources with low optical depth.

There are 20 RSGs in the SMC with DPR values higher than $10^{-9}\, \rm{M_{\odot}\,yr^{-1}}$, contributing to 36\% of the total DPR, while comprising only 0.0097\% of the RSG population. In the LMC, 190 RSGs with DPR values higher than $10^{-9}\, \rm{M_{\odot}\,yr^{-1}}$ contribute to 75\% of the total DPR, comprising  0.04\% of the RSG 
population. The SMC lacks RSGs with high DPR compared to the LMC. Both in the LMC and SMC, a few RSGs play a crucial role in the overall RSGs DPR estimation.
 
The factors influencing the estimation of the DPR have been extensively discussed in detail in  Paper \uppercase\expandafter{\romannumeral1}, we also provided Table~\ref{tab:DPRf} to show the impact of various factors on the estimation of DPR (including AGBs). Because the total DPR depends on a small number of high DPR sources, we have listed in Table~\ref{tab:DPRf} the impact of removed sources in Section~\ref{sec:data} on the result. Although it does have an impact on the cumulative DPR, we believed that these treatments are necessary to ensure the purity of the sample and the accuracy of the fitting. Overall, the DPR is influenced by various factors, such as fitting strategies, model parameter settings, especially depending on the long wavelegth data. Thus the results we presented can be considered as a rough estimation based on current data.
{\catcode`\&=11
\gdef\Dooo {\cite{1995A&A...300..503D}}}
\begin{deluxetable}{c|cc|cc}
\tablecaption{\label{tab:DPRf} Effect of different parameter settings on DPR}
\tablehead{\multicolumn{1}{c}{}&\multicolumn{2}{c}{LMC}&\multicolumn{2}{c}{SMC}\\
\cmidrule(lr){2-3}\cmidrule(lr){4-5}
\colhead{} & \colhead{DPR}&\colhead{$d\dot{M}_{\rm{d}}\%^{a}$} & \colhead{DPR}&  \colhead{$d\dot{M}_{\rm{d}}\%$}\\
\colhead{ }&  \colhead{ ($ \rm{M_{\sun}\,{yr^{-1}}}$)}& \colhead{} &  \colhead{ ($ \rm{M_{\sun}\,{yr^{-1}}}$)}& \colhead{}}   
\startdata
        This work & $9.69 \times10^{-06}$ & - & $1.75 \times10^{-06}$ & -  \\
        No visual inspection$^{\rm b}$& $9.82 \times10^{-06}$ & +1.27\% & $1.77 \times10^{-06}$ & +1.24\%  \\ 
        Original fitting results$^{\rm c}$& $1.04 \times10^{-05}$ & +7.79\% & $1.84 \times10^{-06}$ & +5.20\%  \\ 
        Including sources with $sflag=1$$^{\rm d}$& $1.01 \times10^{-05}$ & +3.87\% & $1.78 \times10^{-06}$ & +2.07\%  \\ 
        Including sources with $sflag=1,-1$$^{\rm e}$ & $1.18 \times10^{-05}$	&+21.27\%	&$2.13 \times10^{-06}$ &	+21.78\%  \\ 
        $E(B-V)=0.1$ (0.033 for the SMC) and $R_{\rm V}=3.1$$^{f}$& $1.07 \times10^{-05}$ & +10.53\% & $1.91 \times10^{-06}$ & +9.23\%  \\ \
        scaling $v_{\rm{exp}}$$^{\rm g}$ & $6.22 \times10^{-06}$ & -35.87\% & $9.46 \times10^{-07}$ & -45.84\%  \\ 
        Mixed dust$^{\rm h}$ & $7.93 \times10^{-06}$ & -18.14\% & $1.39 \times10^{-06}$ & -20.62\%  \\ 
         MIPS/Herschel$^{\rm i}$  & $1.19 \times10^{-05}$&	+798.80\%	&$1.03 \times10^{-06}$&	+1163.72\%  \\
         MIPS/Herschel(good fitting)$^{\rm j}$  & $1.05 \times10^{-06}$ & +51.24\% & - & -  \\ 
\enddata
\tablecomments{``Original fitting results'', ``$E(B - V ) = 0.1$'' and ``Mixed dust'' are original fitting  without restrictions, which are described in Section~\ref{sec:fitting}.  $^{\rm a}$The change rate of DPR compared to ``This work''. $^{\rm b}$Results without visual inspection. $^{\rm c}$Original results obtained based on the minimum $\chi^2$. $^{\rm d}$Results including sources with $sflag=1$, which indicates sources removed based on Simbad. $^{\rm e}$Results including sources with $sflag=1$ and $-1$. $sflag= -1$ indicates sources removed based on data quality.
$^{\rm f}$Results obtained by changing the extinction correction method. 
$^{\rm g}$Results obtained by scaling the wind speed based on \citet{2006ASPC..353..211V}. $^{\rm h}$Results obtained by adding a set of mixed dust models for re-fitting.  $^{\rm i}$Refitting results for sources with MIPS and Herschel data, including 85 sources in the LMC and 19 in the SMC. The cumulative DPR and change rate here only include these sources. $^{\rm j}$Refitting results for sources with MIPS and Herschel data, including 10 sources in the LMC. The cumulative DPR and change rate here only include the 10 sources.}
\end{deluxetable}

\subsubsection{DPR of AGBs}
\label{subsubsec:AGBDPR}
Figure \ref{fig:agbdpr} shows the histograms of AGBs DPR distributions in the LMC and SMC.  C-rich AGBs seem to have a higher overall DPR than O-rich AGBs, while x-AGBs have the highest DPR. The statistical information for AGBs in the LMC is presented in Table~\ref{tab:DPRlmcagb}. Only 11\% of AGBs in the LMC are Op-thin, contributing merely 0.02\% to the total DPR. Op-thin AGBs are much less common compared to RSGs. The majority of AGBs have non-ignorable circumstellar dust shell and contribute to interstellar dust. When considering only chemical types, x-C and C-rich AGBs account for 43\% of the DPR, while x-O and O-rich AGBs contribute 57\% of the DPR, the contributions between different chemical types are not markedly different. Meanwhile, most x-AGBs are C-rich, with only 23 being O-rich, these few x-O AGBs contribute to 23\% of the DPR, even though x-O constitute only 0.07\% of the AGBs sample. 

The DPR statistics for AGBs in the SMC are shown in Table~\ref{tab:DPRsmcagb}. Op-thin AGBs are more common in the SMC than in the LMC,  the proportion is 16\%, yet their contribution to DPR remains a minimal of 0.04\%. Among the 347 x-AGBs, a few x-O with high DPR contribute 22\% of the total DPR, the contribution of C-rich and x-C AGBs does not constitute an absolute majority. The proportions of two chemical type dust are half-half. In the SMC, the contribution of carbon stars is greater than in the LMC, especially for x-C stars.

We discussed the effects of extinction correction (using the same extinction correction method as \citet{2023A&A...676A..84Y}), fitting strategy, and changes in wind speed (scaling wind speed using \citet{2006ASPC..353..211V} relation) on the estimation of DPR. As  presented in Table~\ref{tab:DPRf}. Overall, the estimation of DPR is influenced by many factors and has notable uncertainty.

\begin{figure*}
    \Centering
	\includegraphics[width=0.5\linewidth]{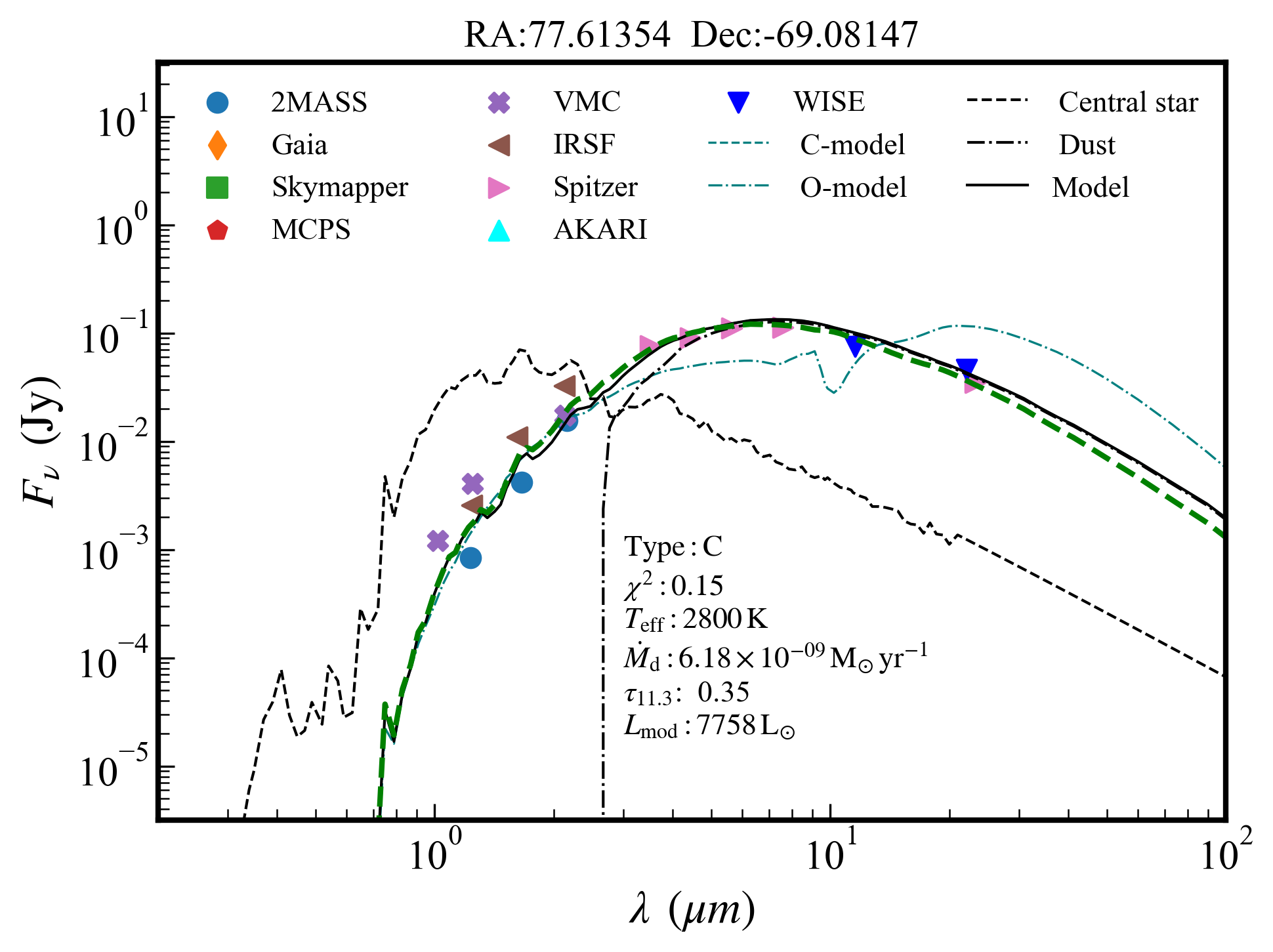}
    \caption{An example of SED fitting with a mixed dust component model. The best-fitting model (black line, C-rich) and mixed dust model (green dash line) are nearly indistinguishable.}
    \label{fig:co}
\end{figure*}

To further discuss the classification of AGBs and DPR, we generated a new set of models with total number over 29,000 
 models. These models' composition include both silicate and amorphous carbon in a ratio of 1:1, and the model mass is in the range of 0.5 to 5 $\rm M_{\sun}$, effective temperature is in the range of 2500 to 5000\,K, other parameters are consistent with  Paper \uppercase\expandafter{\romannumeral1}. We added this new set of models to the model library and refitted the observed SED, 
where the summarized fitting results are also presented in the row of ``mixed dust" of Table~\ref{tab:DPRf}. In cases relying solely on photometric data, it is difficult to distinguish between models with mixed dust components and original models with a single dust component. As shown in Figure~\ref{fig:co}, we present an example where the best-fitting model  and mixed dust model are nearly indistinguishable. 

\begin{figure*}
    \Centering
	\includegraphics[width=1\linewidth]{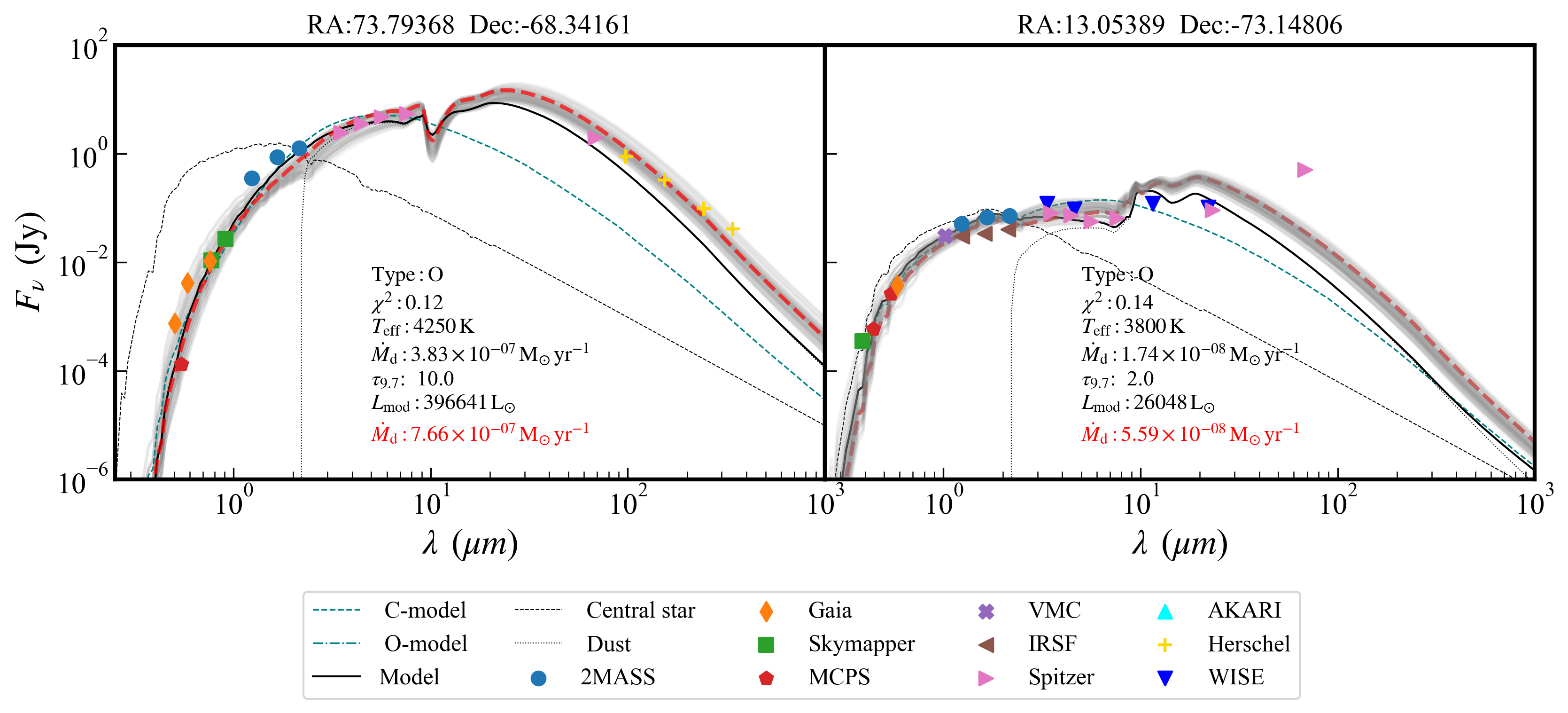}
    \caption{Two examples of refitting after adding MIPS and Herschel data. Only a few sources show reasonable fitting  after the addition of data (left panel, with both MIPS and Herschel data added), while the majority of the refitted results are unreliable (right panel, with only MIPS data added). The solid black line and text represent the original best-fit model and result, while the red line and red text represent the best-fit model obtained from the refitting. The gray lines depict the top 50 models with the minimum $\chi^{2}$ values calculated during the refitting.}
    \label{fig:vz}
\end{figure*}

An accurate estimation of DPR for AGBs requires long-wavelength data. The stellar light of dusty star is largely absorbed by the dust but re-emitted by the dust at $\lambda > 10\,\mu m$，making the circumstellar shell very bright \citep{1999A&A...351..559V}. However, these data is still relatively scarce. We cross-matched our sample with Spitzer MIPS 70$\,\mu$m and 160$\,\mu$m data \citep{2004ApJS..154...25R} and Herschel data \citep{2014AJ....148..124S}, with a radius of 3$^{\prime \prime}$. There were 85 matched sources in the LMC and 19 in the SMC.  We provided two SED fitting examples for these sources in Figure~\ref{fig:vz}. Compared them with the original results in Table~\ref{tab:DPRf},  only a few sources (10 in the LMC and 0 in the SMC) were fitted well (left panel of Figure~\ref{fig:vz}). In the last row of Table~\ref{tab:DPRf}, we listed the DPR and the DPR change rate after changing the fitting for the 10 sources. The 
 additional data didn't improve the accuracy of the fitting. Instead, it had a negative impact in most cases. For example, one of the sources in the SMC had its DPR change from $1.74 \times 10^{-8}\, \rm{M_{\odot}\,yr^{-1}}$ to $5.59 \times 10^{-8}\, \rm{M_{\odot}\,yr^{-1}}$ (right panel of Figure~\ref{fig:vz}), directly changing the statistical results. It can be seen that the estimation of DPR highly depends on the high-quality observational data, especially in the long-wavelength bands.

\subsubsection{Comparison with previous work}

The total DPR of evolved stars in the MCs has been explored by previous studies and compared with our results in Table~\ref{tab:DPRvs}. \citet[hereafter R+12]{2012ApJ...753...71R} estimated the DPR for over 30,000 evolved stars in the LMC using the GRAMS \citep{2011A&A...532A..54S,2011ApJ...728...93S}, while \citet[hereafter B+12]{2012ApJ...748...40B} employed a hybrid approach for estimating the DPR of over 30,000 evolved stars in the MCs based the catalog from \citet{2011AJ....142..103B}. \citet[hereafter S+16]{2016MNRAS.457.2814S} fitted the SEDs of over 4,900 evolved stars in the SMC. 

These studies mainly selected evolved stars based on the SAGE \citep{2006AJ....132.2268M} and SAGE-SMC \citep{2011AJ....142..102G}. 
 Meanwhile, R+21 selected RSGs and AGBs using near-infrared photometry data in the JHK bands from UKIRT/WFCAM and 2MASS, along with Gaia astrometric information. The spatial distribution of our LMC catalog is broader than theirs. However, their SMC catalogs cover the SMC's tail region, which is not covered by our catalog. Another difference is that the selection method of R+21 is based not only on photometric data but also on Gaia astrometric information. For the LMC and SMC, they removed foreground stars based on the proper motion and parallax from Gaia EDR3, which were considered reliable and nearly complete. Therefore, the purity of the sample has been much improved. As mentioned in Section~\ref{sec:data}, we further filtered the sample, and after cross-matching with SIMBAD, we identified only a few contaminating sources.
 
Our total DPR results are similar to the findings of the mentioned studies. We cross-matched with R+12 using radius of $1^{\prime \prime}$, with 29,791 common sources. For these common sources, the total DPR caculated by R+12 is $7.63\times 10^{-6}\, \rm{M_{\odot}\,yr^{-1}}$, while it is $6.31\times 10^{-6}\, \rm{M_{\odot}\,yr^{-1}}$ from our result, which is almost consistent with their results.

R+12 suggested that dust production by C-rich AGBs (including x-C) is 2.5 times that of O-rich AGBs (including x-O), with C-rich AGBs contributing 64.6\% and O-rich AGBs contributing 26.0\% of the total DPR in the LMC, while RSGs contribute 9.4\%. However, our results for the LMC show a slightly lower contribution of x-C and C-rich AGBs (43\%) compared to x-O and O-rich AGBs (57\%). If we only consider sources with $sflag=0$, the contribution of x-C and C-rich AGBs is 47\%, and x-O and O-rich AGBs is 53\%, which are quite close. In the SMC, the contribution of two types AGBs are half-half. When we only consider sources with $sflag=0$, the proportions become 56\% and 44\%. There is no clear difference between the DPR contributions of O-rich and C-rich AGBs,  which is similar to \citet{2013MNRAS.429.2527M}, they utilized MLR-color relations to assess DPR and determined that the distribution of MLR was equal between C-rich and O-rich (O-rich AGBs and RSGs) stars in both the LMC and SMC, and some of the x-AGBs are actually O-rich rather than C-rich. 
Similarly, S+16 used the SED fitting methods to obtain the DPR and found that in the SMC, O-rich (O-rich AGBs and RSGs) and C-rich stars produced dust nearly equal.

As mentioned above, our classification of evolved stars and the estimation of DPR based on photometric data are only rough estimation, as various factors can influence the results obtained. There are often differences in methods and samples between different studies, which may also be a important reason for the discrepancies among them.

\begin{deluxetable*}{c|cc|cc|cc|cc|cc|cc}
\tabletypesize{\tiny}
\tablecaption{\label{tab:DPRvs} Comparison total DPR between our results and previous work}
\tablehead{\multicolumn{1}{c}{}&\multicolumn{2}{c}{This work LMC} & \multicolumn{2}{c}{R+12 LMC $^{a}$} & \multicolumn{2}{c}{B+12 LMC $^{b}$ } & \multicolumn{2}{c}{This work SMC }& \multicolumn{2}{c}{B+12 SMC}& \multicolumn{2}{c}{S+16 SMC $^{c}$}\\
\cmidrule(lr){2-3}\cmidrule(lr){4-5}\cmidrule(lr){6-7}\cmidrule(lr){8-9}\cmidrule(lr){10-11}\cmidrule(lr){12-13}
\colhead{Type}&\colhead{N$^{d}$}&\colhead{Sum} &\colhead{N}&\colhead{Sum}&\colhead{N}&\colhead{Sum}&\colhead{N}&\colhead{Sum}&\colhead{N}&\colhead{Sum}&\colhead{N}&\colhead{Sum}\\
\colhead{}&\colhead{}&\colhead{($\rm{ M_{\sun}\,yr^{-1}}$)} &\colhead{}&\colhead{($\rm{ M_{\sun}\,yr^{-1}}$)}&\colhead{}&\colhead{($\rm{ M_{\sun}\,yr^{-1}}$)}&\colhead{}&\colhead{($\rm{ M_{\sun}\,yr^{-1}}$)}&\colhead{}&\colhead{($\rm{ M_{\sun}\,yr^{-1}}$)}&\colhead{}&\colhead{($\rm{ M_{\sun}\,yr^{-1}}$)}
}
\startdata
        AGB & 30937 & 8.71E-06 & 27615 & 1.91E-05 & 21319 & 1.04E-05 & 5769& 1.64E-06 & 4966 & 8.25E-07 & 3561 & 8.49E-07  \\
        RSG & 4714 & 9.78E-07 &  5876 & 2.00E-06 & 3908 & 2.40E-07 &2057&1.08E-07& 2611 & 3.10E-08 & 1410 & 4.60E-08  \\ 
        Total($sflag=0$ or no FIR) $^{e}$  & 33298 & 8.81E-06 &  - & - & 25406 & 1.07E-05 &7327&1.54E-06& 7577 & 9.50E-07 & 4926 & 8.90E-07  \\ 
        Total & 35651 & 9.69E-06 &  33491 & 2.11E-05 & 25389 & 1.41E-05 &  7825& 1.75E-06&7627 & 8.60E-07 & 4943 & 1.30E-06  \\ 
\enddata
\tablecomments{FIR objects: defined as sources with flux at [24] brighter than at [8.0]. The FIR objects are may potentially include contaminations. $^{a}$\citet{2012ApJ...753...71R}. $^{b}$\citet{2012ApJ...748...40B} $^{c}$\citet{2016ApJ...826...44S}. $^{d}$ Number of sources.}

\end{deluxetable*}

\subsection{Relations between MLR and stellar parameters}

\subsubsection{MLR-stellar parameter relations of RSGs}
MLR can be obtained by multiplying DPR by the gas-to-dust ratio. The gas-to-dust ratio for O-rich AGBs and all RSGs was assumed to be $\phi=1000$ in the SMC and $\phi=500$ in the LMC, for C-rich stars, the ratio was assumed to be $\phi=200$ in both the SMC and the LMC \citep{1999A&A...351..559V,2012ApJ...748...40B,2012ApJ...753...71R}.
Numerous investigations have explored the relation between stellar parameters and the MLR of evolved stars \citep{1975MSRSL...8..369R, 1988A&AS...72..259D, 1992iesh.conf...18F, 1998A&ARv...9...63V, 2005A&A...438..273V, 2017MNRAS.465..403G,2020MNRAS.492.5994B, 2021ApJ...912..112W, 2023A&A...676A..84Y,2023MNRAS.524.2460B,2024arXiv240115163A}. Variations in samples and methods among previous research may lead to considerable dispersion in the obtained MLR prescriptions. In  Paper \uppercase\expandafter{\romannumeral1}, we investigated the relation between the MLR and luminosity of RSGs in the LMC, and found a prominent turning point on the luminosity-MLR diagram. The same method was used to study RSGs in the SMC, with the results presented below. We used third-order polynomials and piecewise functions for fitting, respectively.
The third-order polynomialsm for the SMC is,
\begin{alignat}{4}   
    \log{\dot{M}}&= 0.38\times [\log{(L_{\rm{obs}}/L_{\sun})}]^{3} &&- 4.09\times[\log{(L_{\rm{obs}}/L_{\sun})}]^{2} 
    &&+ 14.65\times[\log{(L_{\rm{obs}}/L_{\sun})}] &&- 25.00 ,\\
    \log{\dot{M}}&= 0.34\times[\log{(L_{\rm{mod}}/L_{\sun})}]^{3} &&- 3.75\times[\log{(L_{\rm{mod}}/L_{\sun})}]^{2} 
    &&+ 13.93\times[\log{(L_{\rm{mod}}/L_{\sun})}] &&-25.00.
\end{alignat}
The piecewise function for the SMC is,
\begin{alignat}{3}
    \log{\dot{M}}&= 0.20\times[\log{(L_{\rm{obs}}/L_{\sun})}] - 8.30\quad&&(\log{(L_{\rm{obs}}/L_{\sun})} <&& 4.52)\notag\\
    &= 2.03\times[\log{(L_{\rm{obs}}/L_{\sun})}] - 16.57\quad  &&(\log{(L_{\rm{obs}}/L_{\sun})} \geq&& 4.52),
\end{alignat}
\begin{alignat}{3}
    \log{\dot{M}}&=0.15\times[\log{(L_{\rm{mod}}/L_{\sun})}] - 8.07\quad&&(\log{(L_{\rm{mod}}/L_{\sun})} <&& 4.34)\notag\\
   &= 1.51\times[\log{(L_{\rm{mod}}/L_{\sun})}] - 13.98\quad&&(\log{(L_{\rm{mod}}/L_{\sun})} \geq&& 4.34).
\end{alignat}

\begin{figure*}
	\includegraphics[width=0.9\linewidth]{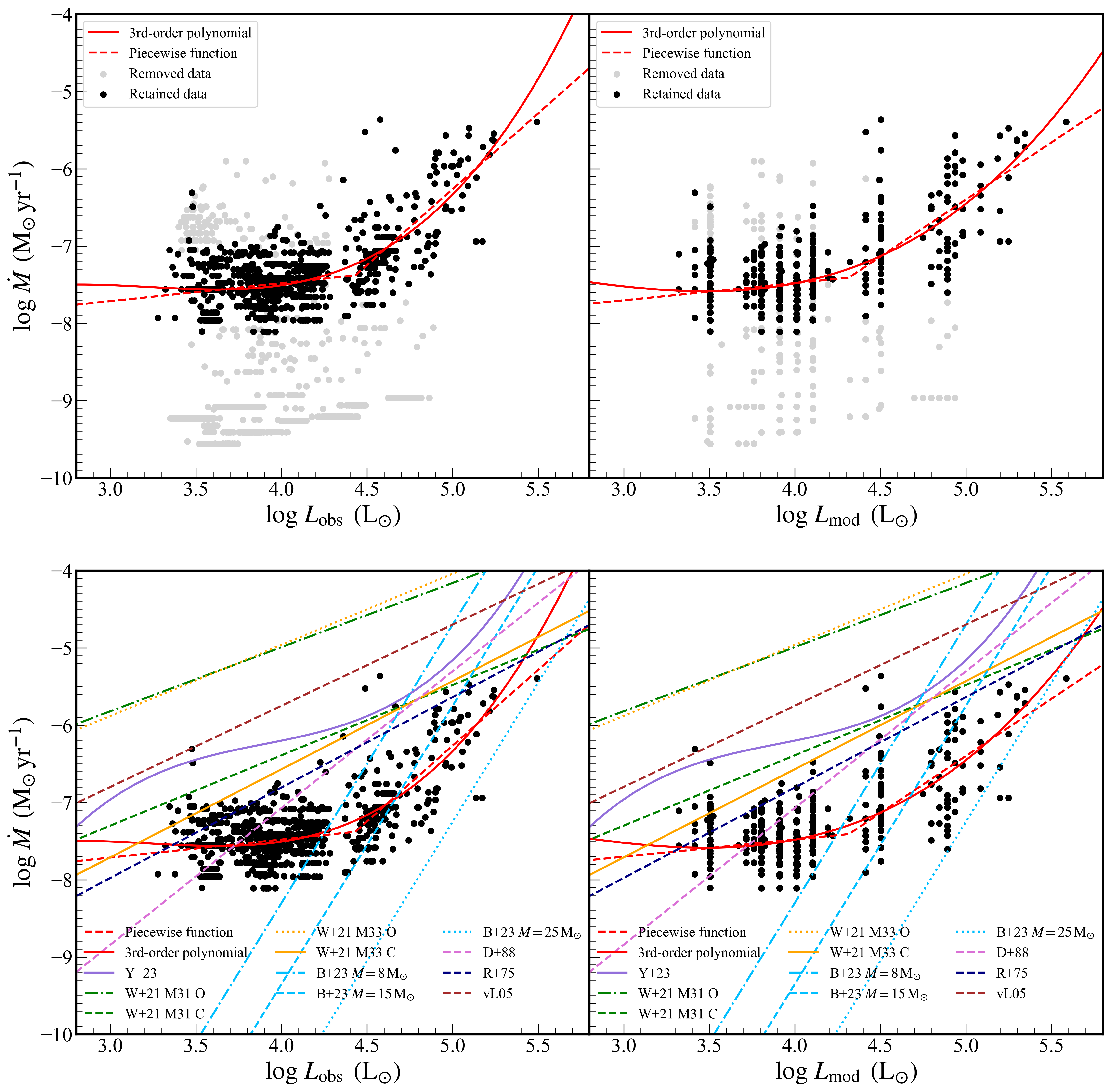}
\caption{Luminosity-MLR diagram of RSGs in the SMC. In the left panel, the luminosity is derived by integrating the observed SED ($L_{\rm{obs}}$), while the right panel depicts the luminosity obtained by integrating the best-fit model SED ($L_{\rm{mod}}$). Excluded stars ($sflag\neq 0$ and $\tau_{\rm{9.7/11.3}} \leq 0.001$) are marked in gray, while retained stars are marked in black. The red curves represent our fitted relations. Bottom row show the comparisons between our luminosity-MLR relations and those from previous studies \citep{1975MSRSL...8..369R,1988A&AS...72..259D,2021ApJ...912..112W,2023A&A...676A..84Y,2023MNRAS.524.2460B,2005A&A...438..273V}.  Various curves represent the relations derived from our work and previous studies.}
    \label{fig:fitrsgsmc}
\end{figure*}
We found turning points of RSG population similar to those in  Paper \uppercase\expandafter{\romannumeral1} and \citet{2023A&A...676A..84Y} as shown in Figure~\ref{fig:fitrsgsmc}. Based on the new model and data presented in this work, we also updated the luminosity-MLR relations of RSGs in the LMC, as shown below.
The third-order polynomialsm for the LMC is,
\begin{alignat}{4}   
    \log{\dot{M}}&= 0.42\times [\log{(L_{\rm{obs}}/L_{\sun})}]^{3} &&- 4.30\times[\log{(L_{\rm{obs}}/L_{\sun})}]^{2} 
    &&+ 14.84\times[\log{(L_{\rm{obs}}/L_{\sun})}] &&- 25.00,\\   
    \log{\dot{M}}&= 0.38\times[\log{(L_{\rm{mod}}/L_{\sun})}]^{3} &&- 4.00\times[\log{(L_{\rm{mod}}/L_{\sun})}]^{2} 
    &&+ 14.22\times[\log{(L_{\rm{mod}}/L_{\sun})}] &&-25.00.
\end{alignat}
The piecewise function for the LMC is,
\begin{alignat}{3}
    \log{\dot{M}}&= 0.22\times[\log{(L_{\rm{obs}}/L_{\sun})}] - 8.65\quad&&(\log{(L_{\rm{obs}}/L_{\sun})} <&& 4.41)\notag\\
    &= 2.57\times[\log{(L_{\rm{obs}}/L_{\sun})}] - 19.02\quad  &&(\log{(L_{\rm{obs}}/L_{\sun})} \geq&& 4.41),
\end{alignat}
\begin{alignat}{3}
    \log{\dot{M}}&=0.23\times[\log{(L_{\rm{mod}}/L_{\sun})}] - 8.67\quad&&(\log{(L_{\rm{mod}}/L_{\sun})} <&& 4.34)\notag\\
   &= 2.23\times[\log{(L_{\rm{mod}}/L_{\sun})}] - 17.36\quad&&(\log{(L_{\rm{mod}}/L_{\sun})} \geq&& 4.34).
\end{alignat}

We conducted a comparison between our results and previous research. \citet{1975MSRSL...8..369R} derived the first widely used relation, which was based on a small sample of stars including both red giants and RSGs. Their relation could be simplified for an average temperature
of $T_{\rm eff} = 3750$\,K \citep[see Equation 14 of][]{2023A&A...676A..84Y}. \citet{1988A&AS...72..259D} collected MLR data for 271 stars with spectral types ranging from O to M. They represented MLR as a function of the effective temperature and luminosity. \citet{2005A&A...438..273V} obtained the MLR formula for O-rich dust-enshrouded AGBs and RSGs in the LMC, which was also  a function of temperature and luminosity. Recently, \citet{2020MNRAS.492.5994B,2023MNRAS.524.2460B} calculated MLR and luminosities for RSGs in clusters in the LMC and derived a prescription related to the initial mass of the stars. \citet{2021ApJ...912..112W} calculated luminosity-MLR relations of RSGs in M31 and M33 based on a large number of RSG candidates from R+21. The relations they obtained can be clearly distinguished into two types: C-rich and O-rich. \citet{2023A&A...676A..84Y} derived the luminosity-MLR formula applicable to RSGs in the SMC, expressed in the form of a third-order polynomial.

The differences in samples, method and model across various studies can result in significant variations among their prescriptions. Formulations derived from small, dust-rich, and bright sources naturally tend to show higher MLR. For studies with larger samples, the environment (e.g. metallicity) is also an important influencing factor. The result of \citet{2021ApJ...912..112W} from the higher metallicity environments of M31 and M33 clearly exceeds the prescriptions we obtained. Consistent with the findings in Paper \uppercase\expandafter{\romannumeral1}, our result closely align with the result reported by \citet{2023MNRAS.524.2460B} for $\log{(L/L_{\sun})} > 4.4$. \citet{2024arXiv240115163A} recent work on RSGs in the LMC  also obtained similar results as Paper \uppercase\expandafter{\romannumeral1}, with a turning point appearing in $\log{(L/L_{\sun})} \approx 4.4$. When  $\log{(L/L_{\sun})} > 4.4$, the luminosity-MLR relation for RSGs becomes more sensitive, suggesting that the mass loss in this region  may be more influenced by radiation pressure. Meanwhile, as revealed by our results in both the LMC and SMC, the proportion of Op-thin RSGs is very high, and the MLR for the majority of RSGs are relatively small in MCs, which is consistent with the findings of \citet{2020MNRAS.492.5994B}. Higher MLR tend to be present in those RSGs with $\log{(L/L_{\sun})} > 4.4$, indicating that mass loss prescriptions derived from a few bright RSGs may not be applicable to all RSGs or the entire RSG phase.

\begin{figure*}
	\includegraphics[width=\linewidth]{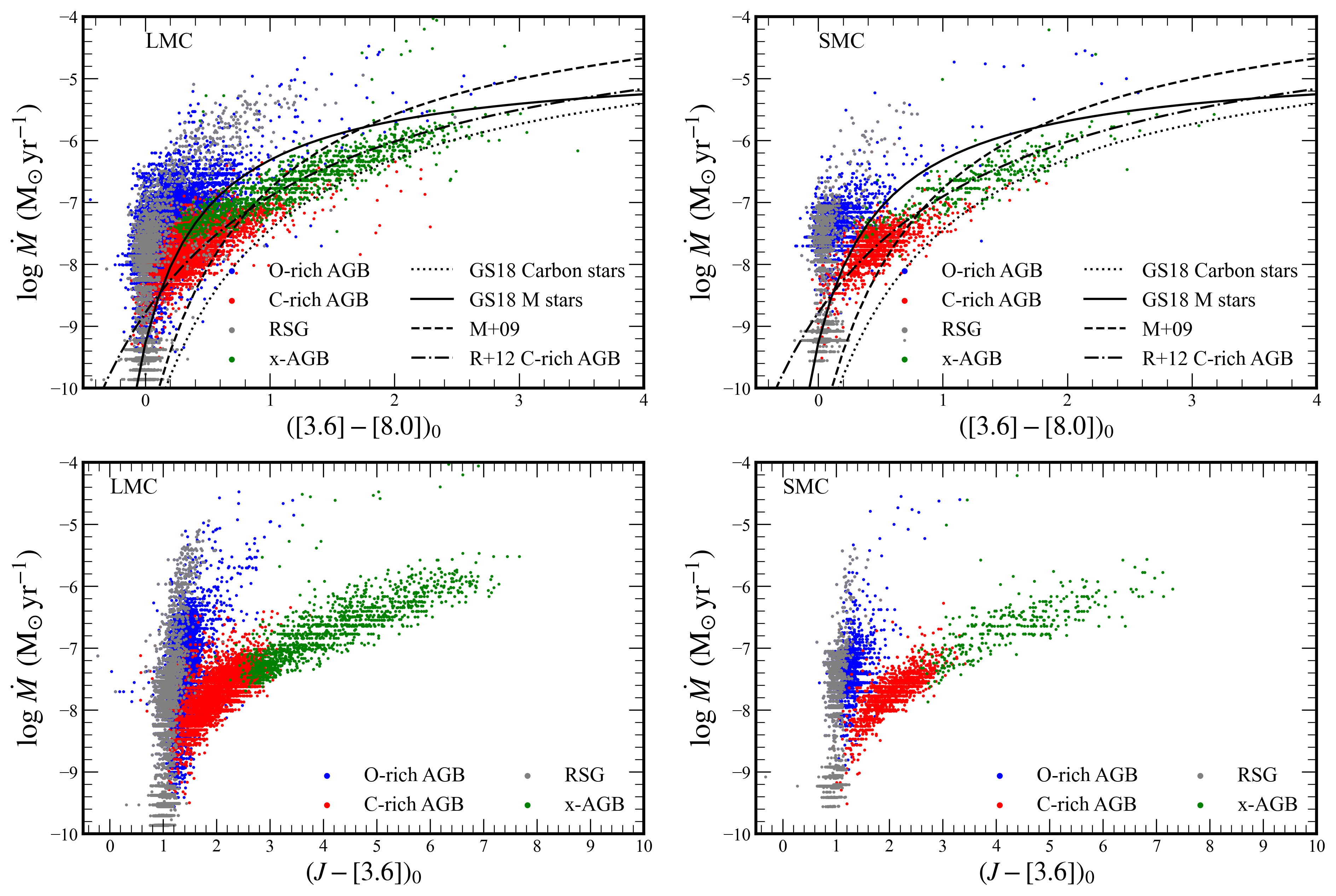}
\caption{ $(\rm{[3.6]- [8.0]})_{0}$ and $(\rm{J - [3.6]})_{0}$ color versus MLR diagrams for the RSGs (gray dots), O-rich AGBs (blue dots), C-rich AGBs (red dots), x-AGBs (green dots). Only sources with $sflag=0$ and not opt-thin are included. The color-DPR relation obtained from previous work is represented by black lines \citep{2009MNRAS.396..918M,2018A&A...609A.114G,2012ApJ...753...71R}.}
    \label{fig:cc2}
\end{figure*}

Another extensively studied prescription is the correlation between MLR and infrared color. \citet{2009MNRAS.396..918M} analyzed AGBs in the LMC and presented the MLR of AGBs as a three-parameter function based on the inverse of the $[3.6]-[8.0]$ color. This function was also used by \citet{2018A&A...609A.114G}, who calculated the MLR and luminosity of 225 carbon stars and 171 O-rich evolved stars in the Local Group. We presented our results as shown in  Figure~\ref{fig:cc2}. A clear positive correlation is observed between RSGs and infrared color, while the relations between different types of evolved stars and infrared color also exhibits noticeable distinctions, RSGs and C-rich AGBs can be well distinguished, but the distribution of low-luminosity RSGs resembles that of O-rich AGBs on the luminosity-MLR diagram and color-MLR diagram. This indicates the difficulty in distinguishing between them 
for which some contamination is inevitable. If we consider only sources with luminosities greater than $10^{4}\,L_{\sun}$ as pure RSGs, then the DPR of RSGs decreases by 12\% in the LMC and 29\% in the SMC.

\subsubsection{MLR-stellar parameter relations of AGBs}

Despite the lack of a comprehensive and accurate theory, many empirical and semi-empirical mass-loss prescriptions have been explored for AGBs. These prescriptions for various phase of AGB evolution, based on stellar parameters such as mass, effective temperature, and luminosity, are widely used in stellar evolution models. Some of the most utilized and recent prescription include those by \citet{1975MSRSL...8..369R,1993ApJ...413..641V,1995A&A...297..727B,2002A&A...384..452W,2005A&A...438..273V,2005ApJ...630L..73S}, as well as those based on dynamical model atmospheres by \citet{2010A&A...509A..14M,2014A&A...566A..95E} for C-stars, and \citet{2019A&A...623A.119B} for M-type AGB stars.  The simple luminosity-MLR for different types of AGBs were fitted using different linear functions as,
\begin{equation}
\label{eq:ab}
    \log{\dot{M}}=a\log{L_{\rm{mod/obs}}}+b
\end{equation}

The fitting results are illustrated in Figure~\ref{fig:fitdatalmcagb} and \ref{fig:fitdatasmcagb}, where only sources with $sflag=0$ and $\tau_{\rm{9.7/11.3}} > 0.001$ were adopted in the fitting process. The coefficients obtained from the fitting are listed in Table~\ref{tab:ab}. In Figure~\ref{fig:fitdatalmcagb} and ~\ref{fig:fitdatasmcagb}, there is a positive correlation between the MLR and luminosity. Overall, the distribution of AGBs on the luminosity-MLR diagram is so extensive that determining a comprehensive luminosity-MLR relation is almost meaningless, highlighting the complexity of the AGBs mass-loss mechanism. The spread is partly attributed to differences in the evolutionary phase. A simple luminosity-dependent MLR prescription can't apply to the entire evolutionary phase of the AGBs. However, we believe that accuracy in the luminosity-MLR relations can be improved by classifications of AGBs. 

\begin{figure*}
	\includegraphics[width=0.8\linewidth]{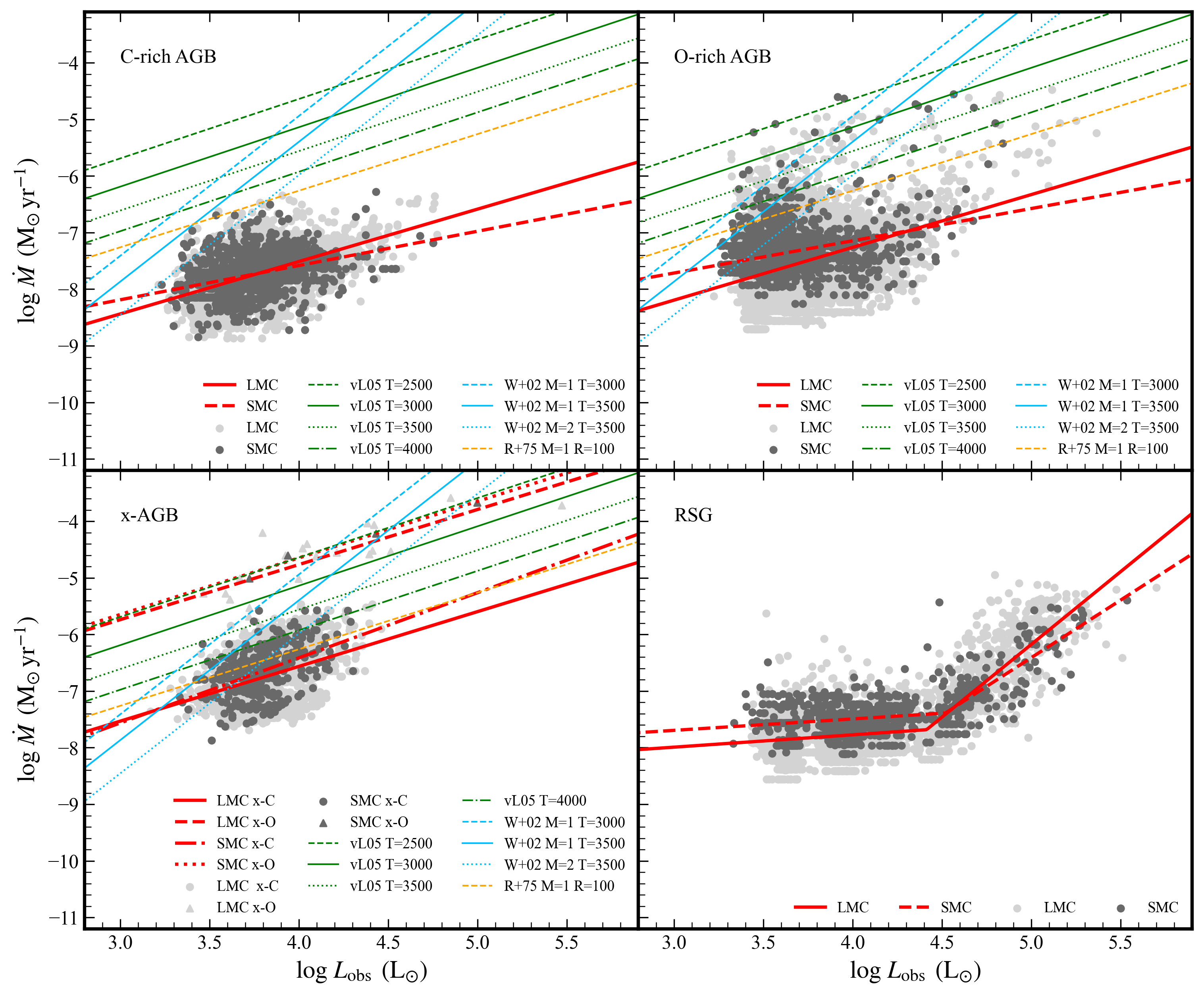}
    \centering
\caption{Luminosity-MLR diagrams for different types of evolved stars in MCs, where light gray represents sources in the LMC, dark gray represents sources in the SMC. The red lines represent our fitted prescription. The lines with various colors represent the prescriptions derived from previous studies \citep{1975MSRSL...8..369R,2002A&A...384..452W,2005A&A...438..273V}.}
    \label{fig:mcs}
\end{figure*}

\begin{figure*}
	\includegraphics[width=\linewidth]{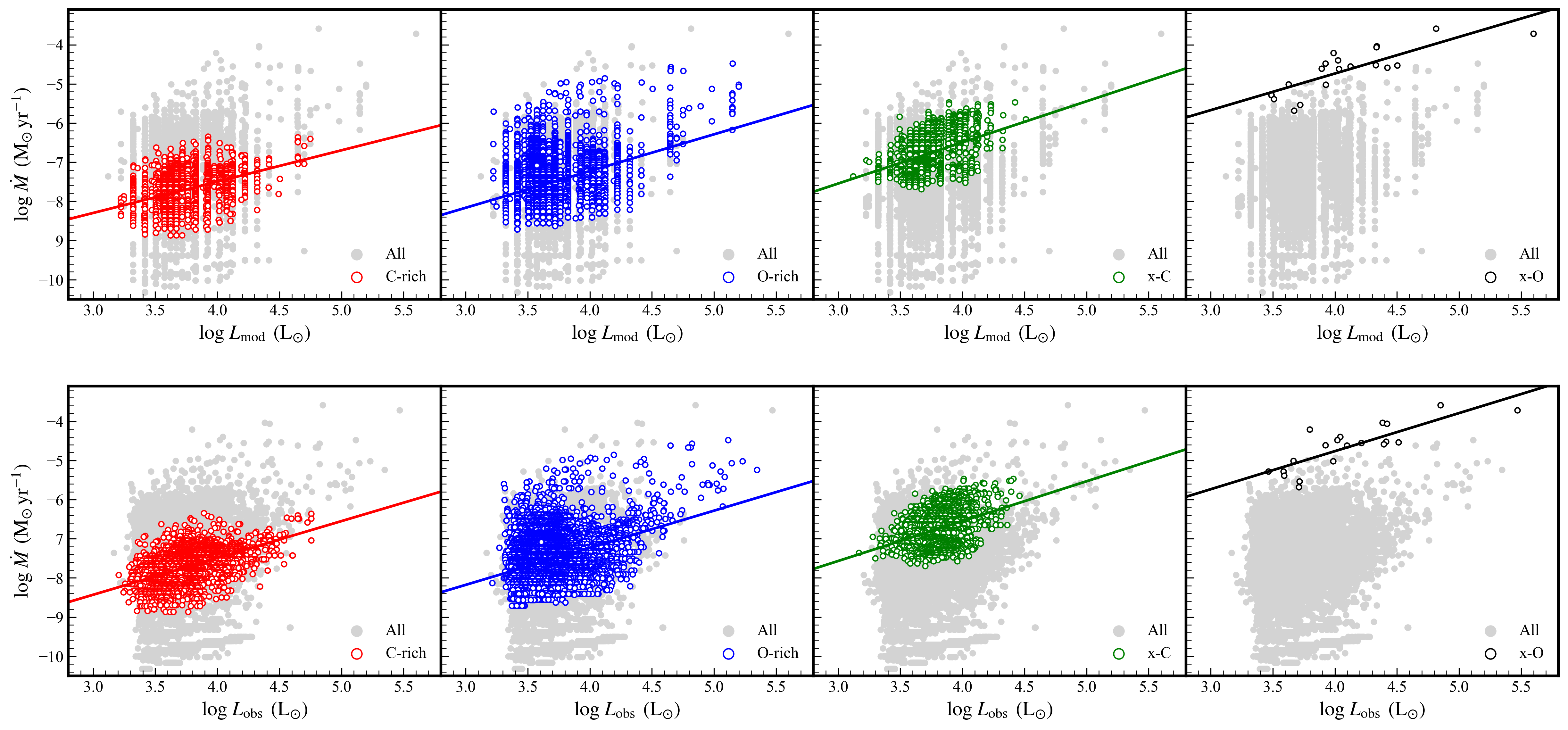}
\caption{Luminosity-MLR diagrams for different types of AGBs in the LMC. The top panel shows the luminosity derived from integrating the best-fit model SED ($L_{\rm{mod}}$), while the bottom panel shows the luminosity derived from integrating the observed SED ($L_{\rm{obs}}$)}
    \label{fig:fitdatalmcagb}
\end{figure*}

\begin{figure*}
	\includegraphics[width=\linewidth]{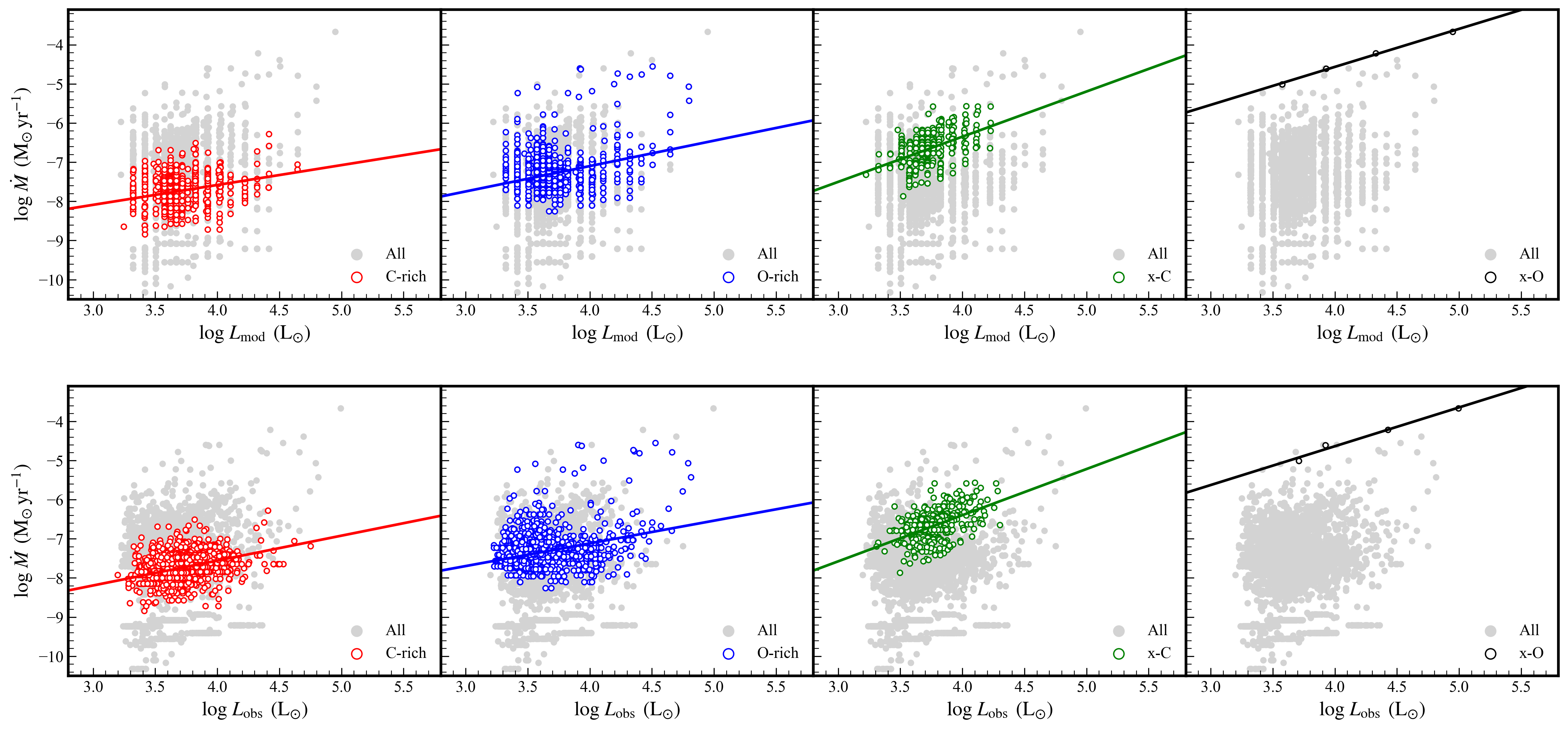}
\caption{Same as Figure~\ref{fig:fitdatalmcagb} but for AGBs in the SMC.}
    \label{fig:fitdatasmcagb}
\end{figure*}

Figure~\ref{fig:mcs} shows a comparison between our results and the prescriptions obtained by previous work. \citet{2005A&A...438..273V} derived a formula valid for dust-enshrouded RSGs and O-rich AGB stars, which depends on temperature and luminosity. Their formula matchs well with the scatter of x-O and a small portion of O-rich AGBs with high DPR,  which also almost coincides with ours formula for x-O stars. However, for most O-rich AGBs, this formula may overestimate the DPR. For prescriptions derived by \citet{2002A&A...384..452W}, the situation is similar, as these prescriptions are more suitable for a small number of evolved stars.

The MLR (DPR)-infrared color has also been widely studied and utilized. As shown in Figure~\ref{fig:cc2}, there is a clear positive correlation between  DPR and infrared colors, O-rich and C-rich stars can also be well distinguished. The Figure~\ref{fig:cc2} shows that our results for C-rich and x-C AGBs are highly consistent with R+12. We found that both in $([3.6]-[8.0])_{0}$ and $(J-[3.6])_{0}$ diagrams, C-rich stars and O- rich stars can be separated more clearly in the SMC.

\begin{deluxetable*}{c|cccc|cccc}
\tabletypesize{\small}
\tablecaption{\label{tab:ab} The coefficient of the luminosity-MLR relations of AGBs}
\tablehead{\multicolumn{1}{c}{}&\multicolumn{4}{c}{LMC}&\multicolumn{4}{c}{SMC}\\\cmidrule(lr){2-5}\cmidrule(lr){6-9}
\colhead{~} & \colhead{C-rich} & \colhead{O-rich} & \colhead{x-C} & \colhead{x-O} & \colhead{C-rich} & \colhead{O-rich} &\colhead{x-C} & \colhead{x-O}} 
\startdata
        a(mod) & 0.80 &	0.94 &	1.05 &	0.94 	&0.51 &	0.65 &	1.15 &	0.97  \\ 
        b(mod) & -10.70 &-10.98 &-10.70 &-8.48 	&-9.61 &-9.69 &	-10.96 &-8.45  \\  
        a(obs) & 0.93 	&0.93 &	0.97 &	0.97 &	0.61 &	0.57 &	1.15 &	1.00  \\ 
        b(obs) & -11.21 &-10.99 &-10.43 &-8.66& -10.00 &	-9.42 &	-11.01 &	-8.65 
  \\
\enddata
\end{deluxetable*}

\begin{figure*}
\centering
	\includegraphics[width=1\linewidth]{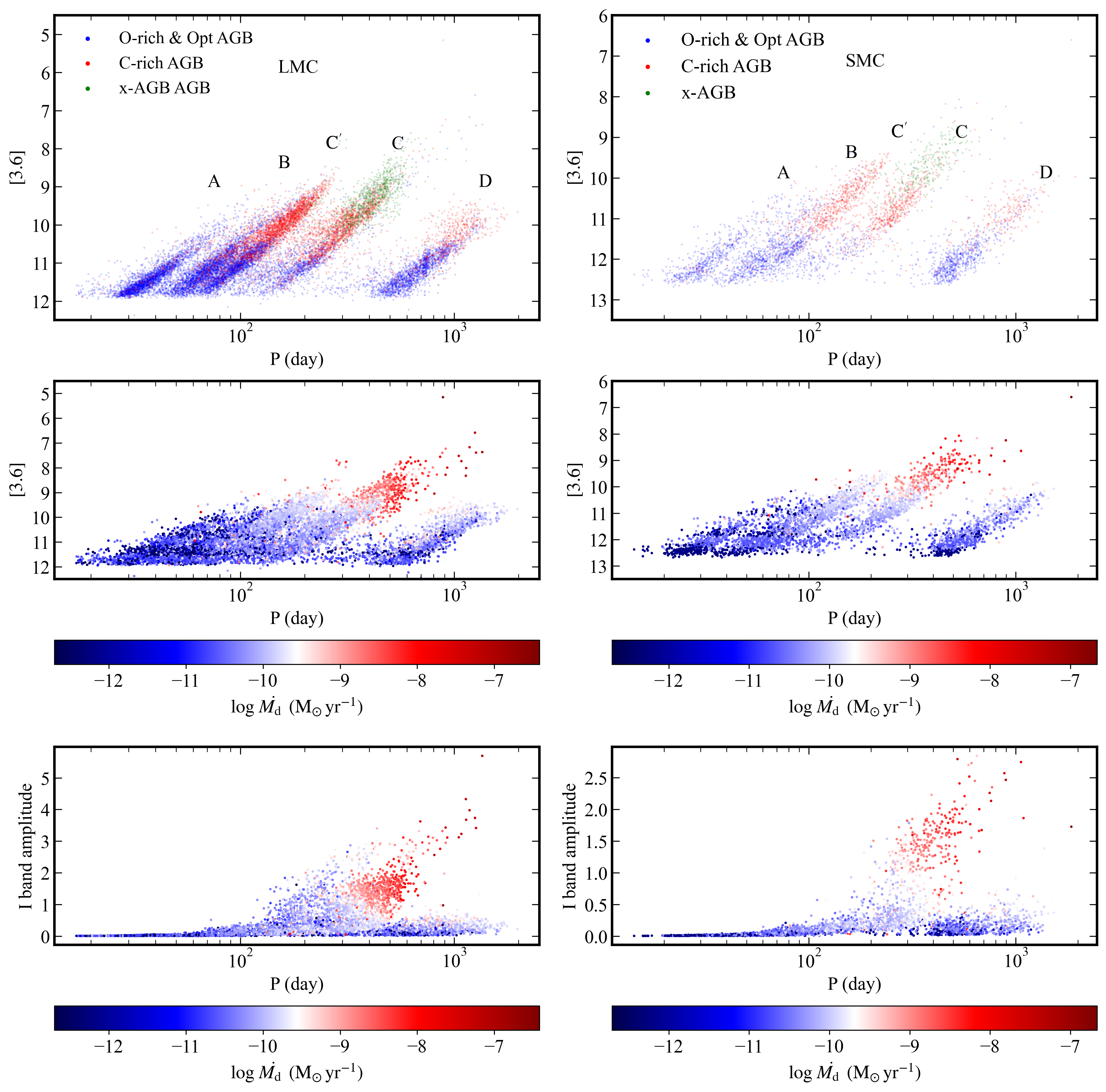}
\caption{The diagrams depicting the relations between DPR and pulsation parameters for AGBs ($sflag=0$) in the LMC (left) and SMC (right). The first row illustrates the changes in AGBs types with the evolution of pulsation sequences. The second row shows the change of DPR on the period-[3.6] diagram. The third row shows the variation of DPR on the period-amplitude diagram. }
        \label{fig:PAdpr}
\end{figure*}

\begin{figure*}
\centering
	\includegraphics[width=0.7\linewidth]{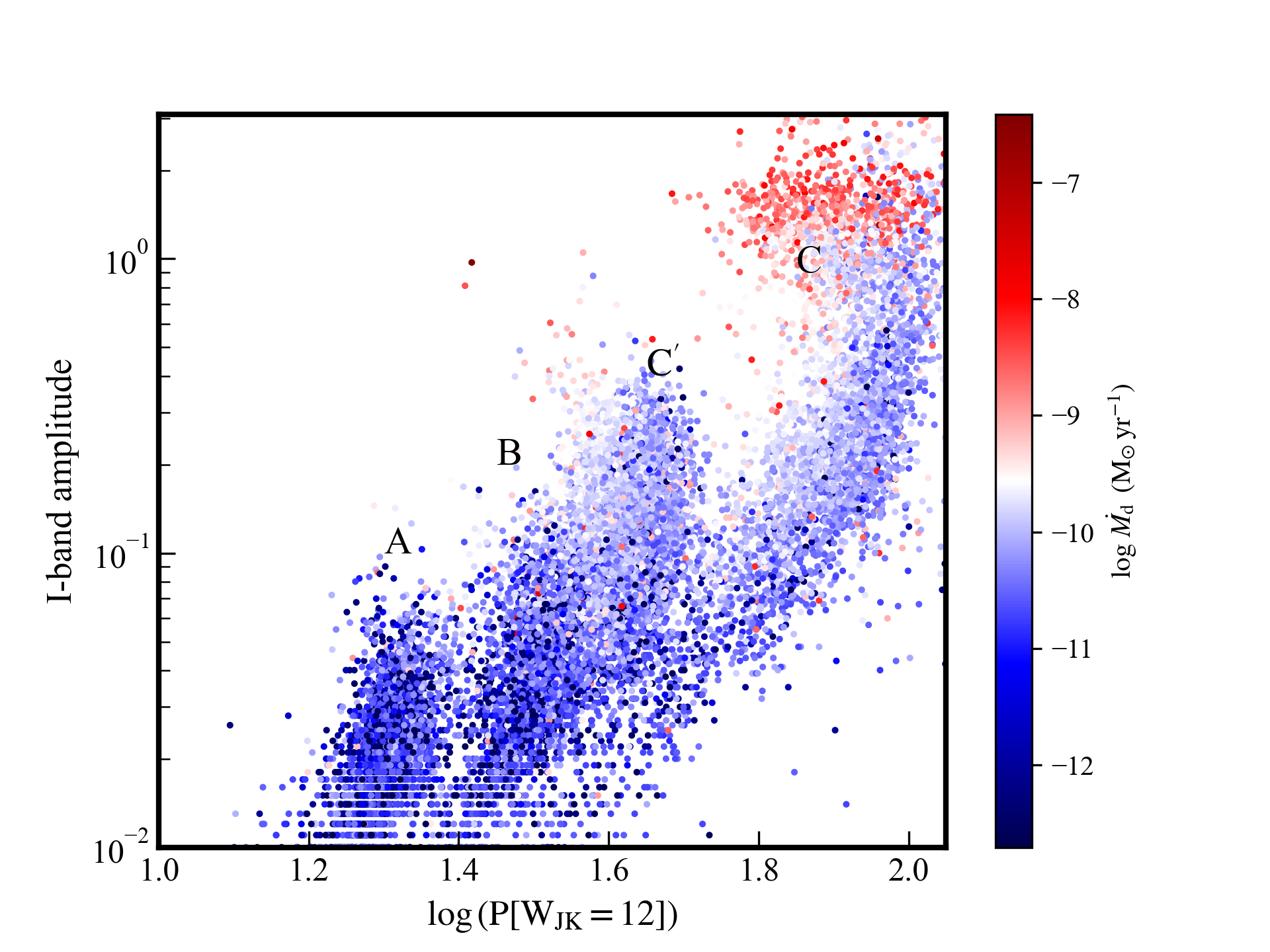}
\caption{Projected pulsation period versus amplitude diagram. The period is replaced with the parameter \rm${log(P[W_{JK}=12])}$.}
    \label{fig:d2}
\end{figure*}

A notable feature of AGBs is the variability caused by pulsations. Pulsation is also considered a crucial mechanism for the mass-loss in AGBs, with strong mass-loss is likely initiated by pulsations. Properties of variable stars are often traced through the period-luminosity diagram \citep[e.g.][]{2010ApJ...723.1195R,2015MNRAS.448.3829W}. To achieve this, we cross-matched our AGBs sample with the Optical Gravitational Lensing Experiment (OGLE) Catalogue of Long-Period Variables \citep{2009AcA....59..239S,2011AcA....61..217S}, using a cross-radius of 1$^{\prime \prime}$. There are 20,496 common sources in the LMC and 4,716 in the SMC, respectively. Figure~\ref{fig:PAdpr} illustrates the relation between DPR of AGBs (only with $sflag=0$) and pulsation parameters. The first row shows the changes in AGBs types with the evolution of pulsation sequences, from left to right as sequences A to C and D. During the evolution from sequence A to C, the proportion of C-rich AGBs increases. In the sequence C, x-AGBs start to appear when the period reach approximately 300 days. The second row shows the variation of DPR on the period-[3.6] diagram. The DPR of AGBs change with the evolution of sequences A to C, and a significant increase occurs when the period reaches 300 days. This is consist with previous studies \citep[e.g.][]{1977ApJ...211..499W,2000A&A...361..641W, 2016ApJ...823L..38M}. The third row shows the variation of DPR on the period-amplitude diagram. AGBs show a noticeable increase in DPR as the amplitude increases, especially when the I band amplitude is greater than 0.5\,mag. Moreover, \citet{2019MNRAS.484.4678M} found a distinct transition in the period-amplitude diagram from sequence B to $\rm{C}^{\prime}$ (see panel(b) of Figure~2 in their work). Using the same projection method, which was described in \citet{2019MNRAS.484.4678M} and \citet{2015MNRAS.448.3829W}, we found that their conclusion could be reproduced well in our result, as shown in Figure~\ref{fig:d2}. A distinct transition of DPR appears at $\rm{log(P[W_{JK}=12])}$ $\sim$ 1.55\,dex ($\rm{log(P[W_{JK}=12])}=\log{P} + (W_{\rm JK}-12)/4.444$), indicating that pulsation is a key factor to initiate the mass-loss of AGBs.

\subsection{LMC and SMC}

In this work, the estimation of DPR and stellar parameter uncertainties introduce a high degree of uncertainty, for which tracing the effects of metallicity can be challenging. 

 Previous works have found some dependence of MLR or DPR on metallicity \citep[e.g.][]{2009A&A...506.1277G,2016ApJ...826...44S,Sloan_2012,Boyer_2015,Boyer_20152,2012ApJ...752..140S,2010ApJ...719.1274S}.  In our results, the main differences between the LMC and SMC are primarily evident in the proportion of Op-thin evolved stars. In the SMC, the proportions of Op-thin RSGs and AGBs are 53\% and 16\%, respectively, higher than that in the LMC (39\% and 11\%). Another noticeable difference is that RSGs contribute approximately 13\% to the total dust budget of the LMC, which is higher than 6\% in the SMC, and the average DPR of RSGs in the LMC is 3.98 times that of the SMC. Additionally, the proportion of RSGs in the LMC with DPR exceeding $10^{-9}\, \rm{M_{\odot}\,yr^{-1}}$ (0.04\%) is higher than that in the SMC (0.0097\%). These finding aligns with \citet{2000A&A...354..125V,2006ASPC..353..211V} and \citet{2008ApJ...686.1056S}, indicating that the formation of O-rich dust is more challenging under conditions of low metallicity, with O-rich stars DPR seems to decrease with decreasing metallicity \citep{2009A&A...506.1277G}.  As shown in Figure~\ref{fig:mcs}, another distinction between the LMC and SMC is reflected in the distribution of RSGs in the luminosity-MLR diagram. The slope of RSGs after the tuning point in the SMC appears to be less steep. There is a noticeable difference in the distribution of effective  temperatures and luminosity for the evolved stars of the LMC and SMC, with the overall effective temperature in the SMC being slightly higher than that in the LMC by approximately 100\,K (see Figure~\ref{fig:tefflumtau}), while the luminosity is opposite.

Some previous studies suggested that at low metallicity, the DPR of carbon stars would be lower \citep{2000A&A...354..125V,2006ASPC..353..211V,2008A&A...487.1055V}. \citet{2012ApJ...748...40B} found that the average DPR of carbon stars in the LMC is 2.2 to 4.9 times higher than that in the SMC. However, we found that the DPR of C-rich AGBs in the SMC is just slightly lower than in the LMC, with values of $1.26 \times 10^{-10}\, \rm{M_{\odot}\,yr^{-1}}$ and $1.15 \times 10^{-10}\, \rm{M_{\odot}\,yr^{-1}}$, respectively. The difference between them is marginal. For x-C, there is almost no difference in the average DPR between the SMC and LMC, similar to the conclusion drawn by \citet{2007MNRAS.376..313G}, \citet{2008ApJ...686.1056S}, and \citet{2007MNRAS.382.1889M}, where  metallicity dependence for DPR or MLR of carbon stars was not observed. To accurately study the dependence of metallicity, further detailed investigation and more data are needed.

\subsection{Interstellar dust}
 Evolved stars are considered one of the most important sources of interstellar dust. Previous studies have estimated the total dust mass in the LMC and SMC using different methods. \citet{2010A&A...523A..20B} obtained total dust mass of  $3.6 \times 10^{6}\,\rm{M_{\odot}}$ and $0.29-1.1\times 10^{6}\,\rm{M_{\odot}}$ for the LMC and SMC, respectively. On the other hand, \citet{2014ApJ...797...85G} reported results of approximately  $7.3 \times 10^{5}\,\rm{M_{\odot}}$ and $8.3 \times 10^{4}\,\rm{M_{\odot}}$ for the LMC and SMC, respectively. We can perform a simple estimation by dividing the total dust mass by the DPR, which allows us to estimate the  timescale of dust replenishment from evolved stars. For the total dust mass provided by \citet{2010A&A...523A..20B}, the replenishment timescale in the LMC and SMC is approximately $3.7 \times 10^{11}\, \rm{yr}$ and $1.7-6.2 \times 10^{11}\, \rm{yr}$, respectively. In the case of \citet{2014ApJ...797...85G} , the replenishment timescales are approximately $7.5 \times 10^{10}\, \rm{yr}$ and $4.7 \times 10^{10}\, \rm{yr}$ for the LMC and SMC, respectively.

If we consider the uncertainty in the DPR, the timescale could be extended or shortened by several times.  Of course, the estimation of total dust mass and DPR is fraught with large uncertainty, and factors such as dust lifetime \citep[e.g.][]{2015ApJ...799..158T}, the contribution of supernovae to dust \citep[e.g.]{2006Sci...313..196S,2011Sci...333.1258M}, and their destructive effects are still not well-defined \citep[e.g.][]{2015Sci...348..413L,2012ApJ...748...12S}. Nevertheless, overall, it seems that we need to consider dust sources beyond evolved stars to account for the currently observed dust mass.

\section{Summary}
\label{sec:summary}
Based on a new complete sample of approximately 40,000 evolved stars in the MCs, we adopted an uniform method of SED fitting to systematically investigate the relations between chemical types, DPR, and the stellar parameters, providing a catalog that included stellar and dust envelope parameters.

In our results, the total DPR for RSGs and AGBs in the LMC is approximately $9.69 \times10^{-6}\,\rm{M_{\odot}\, yr^{-1}}$, while it is around $1.75\times10^{-6}\,\rm{M_{\odot }\, yr^{-1}}$ in the SMC, with RSGs DPR accounting for only 13\% and 6\% of the total DPR. There is no significant difference in the contribution of O-rich stars and C-rich stars to the dust. The x-AGBs dominate the dust production, with x-O making a notable contribution. A few evolved stars play a crucial role in the DPR estimation. We find that the calculation of DPR is highly influenced by parameter settings, data quality and other factors. The Op-thin proportion in the SMC is higher than that in the LMC, and the contribution of RSGs to total DPR is lower than that in the LMC, which may be due to the effect of metallicity.

We investigate the relations between stellar parameters and MLR of evolved stars. We fit the luminosity-MLR relation, revealing a turning point at $\log{(L/L_{\sun})} \approx 4.4$ in the luminosity-MLR diagram for RSGs. The luminosity-MLR distribution for AGBs shows obvious dispersion, requiring the fitting based on different types. In general, the MLR of AGBs is not sensitive to changes in luminosity. The MLR of all evolved stars have clear positive relation with infrared colors, and different types of evolved stars can be roughly distinguished in the color-MLR diagram. The DPR of AGBs show a noticeable increase with the evolution of pulsation period and amplitude in the sequence C. When the period approaches 300 days and the I band amplitude is greater than 0.5 mag, DPR increases dramatically. An obvious transition of DPR appear at the middle of sequence B to $\rm{C}^{\prime}$. Pulsation may indeed be a crucial mechanism for AGBs mass loss, while for RSGs, radiation pressure may play a more critical role in mass loss.

Despite the numerous studies on the MLR of evolved stars, differences in samples and methods among various studies lead to widely varying results. Therefore, using a unified method to study a new large sample to obtain statistically significant results is important for understanding of MLR of evolved stars and also beneficial for subsequent research. In this work, we applied the same data screening criteria, processing methods, uniform models, and parameter settings to large samples in MCs, avoiding the impacts caused by differences in samples and parameter settings. This approach may help further investigation on  dust production, mass loss mechanisms, and the role of evolved stars as dust producers, aiding our understanding of mass loss mechanisms and dust production, and also help us better understand the impact of metallicity on dust production.

\section* {}

We appreciate the patient review and suggestions of the anonymous reviewer, which are crucial for us to improve this paper. This work is supported by the National Key R\&D Program of China   No. 2019YFA0405500 and No. 2019YFA0405501, National Natural Science Foundation of China No. 12133002 and U2031209, 12203025, 12173034, 12373048 and 11833006, and Yunnan University grant No. C619300A034, Shandong Provincial Natural Science Foundation through project ZR2022QA064. We acknowledge the science research grants from the China Manned Space Project with No. CMS-CSST2021-A09, CMS-CSST-2021-A08 and CMS-CSST-2021-B03. The numerical computations were conducted on the Yunnan University Astronomy Supercomputer and the Qilu Normal University High Performance Computing, Jinan Key Laboratory of Astronomical Data. This research made use of the cross-match and other service provided by CDS, TOPCAT \citep{2005ASPC..347...29T} and Astropy. 


\bibliography{sample631}{}
\bibliographystyle{aasjournal}


\end{CJK*}
\end{document}